\documentclass[10pt]{iopart}
\usepackage{iopams} 

\expandafter\let\csname equation*\endcsname\relax
\expandafter\let\csname endequation*\endcsname\relax
\usepackage{amsmath}

\usepackage{graphicx}
\usepackage{bm}
\usepackage{etoolbox}
\usepackage{physics}
\usepackage{stackengine}
\usepackage{mathtools}
\usepackage{lipsum}
\usepackage{breqn}
\usepackage{hyperref}
\usepackage{nicefrac}
\usepackage{dblfloatfix}
\usepackage{afterpage}
\usepackage{amsmath,amsfonts}
\usepackage{siunitx}
\usepackage{subfig}
\usepackage{cuted}
\usepackage{cleveref}
\usepackage{cite}
\usepackage{float}
\def\XXint#1#2#3{{\setbox0=\hbox{$#1{#2#3}{\int}$ }
\vcenter{\hbox{$#2#3$ }}\kern-.6\wd0}}

\hypersetup{
    colorlinks=true,
    linkcolor=blue,     
    urlcolor=cyan,
    citecolor=blue,
}

\AtBeginEnvironment{align}{\setcounter{subeqn}{0}}
\newcounter{subeqn} \renewcommand{\thesubeqn}{\theequation\alph{subeqn}}%
\newcommand{\subeqn}{%
  \refstepcounter{subeqn}
  \tag{\thesubeqn}
}

\begin{document}

\title[]{Theoretical description of chirping waves using phase-space waterbags}

\author{H. Hezaveh$^1$, Z. S. Qu$^1$, M. J. Hole$^{1,2}$ and R. L. Dewar$^1$}

\address{$^1$ Mathematical Sciences Institute, The Australian National University}
\address{$^2$ Australian Nuclear Science and Technology Organisation, Locked Bag 2001, Kirrawee DC, NSW, 2232, Australia}

\ead{hooman.hezaveh@anu.edu.au}
\date{today}

\begin{abstract}

The guiding centre dynamics of fast particles can alter the behaviour of energetic particle driven modes with chirping frequencies. In this paper, the applicability of an earlier trapped/passing locus model [H. Hezaveh et al 2017 Nucl. Fusion 57 126010] has been extended to regimes where the wave trapping region can expand and trap ambient particles. This extension allows the study of waves with up-ward and down-ward frequency chirping across the full range of energetic particle orbits. Under the adiabatic approximation, the phase-space of energetic particles is analysed by a Lagrangian contour approach where the islands are discretised using phase-space waterbags. In order to resolve the dynamics during the fast formation of phase-space islands and find an appropriate initialisation for the system, full-scale modelling is implemented using the bump-on-tail (BOT) code. In addition to investigating the evolution of chirping waves with deepening potentials in a single resonance, we choose specific pitch-angle ranges in which higher resonances can have a relatively considerable contribution to the wave-particle interaction. Hence, the model is also solved in a double-resonance scenario where we report on the significant modifications to the behaviour of the chirping waves due to the $2^{\text{nd}}$ resonance. The model presented in this paper gives a comprehensive 1D paradigm of long range frequency chirping signals observed in experiments with both up-ward and down-ward chirping and multiple resonances. 

\end{abstract}

%
%
%
%
\ioptwocol

\section{Introduction}
\label{sec:intro}

The confinement of energetic particles (EPs), which affects the operation of a fusion device, can be markedly modified by their interaction with weakly damped plasma waves \cite{Gorelenkov2014,Gorelenkov2003,Fasoli2007}. In case of inverse Landau damping in a bump-on-tail model (BOT), the nonlinear saturation of the eigenmode \cite {Berk1992,Berk1993,Boris1993} due to particle trapping aligns with flattening of the distribution function of energetic particles \cite{ONeil1965}. During this process, sideband oscillations emerge and if the system provides these oscillations with weak damping, they may develop into signals with chirping frequencies \cite{Lilley2014}. This phenomenon is governed by the fast formation of phase-space islands i.e. the holes and clumps, in the generalised phase-space of energetic particles \cite{Berk1997}. Once formed, these structures evolve slowly in time hence the adiabatic invariant of the EPs trapped in the chirping mode is conserved. In realistic geometries and for long deviations of the frequency from the initial eigenfrequency \cite{Gryaznevich2000,Maslovsky2003,Fredrickson2006}, the EPs can be carried by the wave potential on slices of the phase-space which results in a change in particles toroidal angular momentum \cite{Hezaveh2020,Wang2018}. This is in conjunction with a change in the number of the flux surface on which the particles lie. Consequently, an inward or outward convective transport of the EPs occur leading to unwanted confinement losses. Therefore, it is essential to perform a detailed study of holes and clumps shape as well as EPs dynamics to identify and control the hard nonlinear evolution of an EP driven mode. 

For highly passing EPs, the theoretical picture of long range adiabatic frequency chirping, using a Langmuir wave as an example, was first developed by Breizman \cite{Boris2010}. At each frequency, the nonlinear wave equation is represented as the long-term solution of a Vlasov-Poisson system, hence called a BGK-type mode \cite{BGK}. Subsequently, the impact of EPs collisions, namely Krook, drag and diffusion was studied by Nyqvist \etal in Refs. \cite{Nyqvist2012} and \cite{Nyqvist2013}. The latter allows the separatrix to expand and trap new EPs. Hezaveh \etal \cite{Hezaveh2017} studied the impact of energetic particle orbit topologies on the long range frequency sweeping of a BGK-type mode. This model shows how the inclusion of trapped particle orbits as well as barely passing types can considerably alter the behaviour of a nonlinear chirping wave. For the range of magnetically trapped EPs and a constant trend i.e. up-ward or down-ward in frequency chirping, it has been shown that the trapping region of the BGK mode may initially grow and then shrink (see fig.6 in \cite{Hezaveh2017}). In the topic of long range adiabatic frequency chirping, this model is comprehensive from the perspective of capturing a range of typical guiding centre orbits. However, the assumption of a flat-top phase-space density across the trapping region (separatrix) restricts the applicability of this model only to the regions where the trapping region of the perturbed mode shrinks and particle trapping due to the expansion in phase-space is avoided.  In this work, we aim to relax the flat-top assumption of Ref. \cite{Hezaveh2017} and extend the trapped-passing locus model to cases where the wave potential can deepen and trap new ambient particles as well as shrink leading to a loss of trapped particles. Consequently, this allows us to explore the adiabatic evolution of the chirping wave over the full range of EPs orbits for both up-chirping and down-chirping BGK modes.

In Ref. \cite{Nyqvist2013}, the adiabatic evolution of phase-space holes has been studied in a system where these structures are initialised somewhat off the linear resonance using a given initial profile and a grid-based numerical method.  The claim that holes and clumps form off the initial resonance consists with theory \cite{Lilley2014} and numerical simulations \cite{Lilley2010}. Nevertheless, the initial profile of the clumps is chosen such that the amplitude of the chirping wave is a smooth function of time. This may not necessarily correspond to a proper initial shape for the just-formed phase-space structures. In this regard, a more comprehensive approach is to apply full-scale modelling to the fast formation stage of these structures to find their phase-space profile prior to the adiabatic evolution. Accordingly, we also perform simulations using the BOT code, developed by Lilley \cite{Lilley2010}, and initialise the phase-space using the simulation data. Subsequently, we resolve the EPs response to the chirping mode using a non-perturbative approach under the adiabatic ordering. For a growing separatrix in phase-space, we implement a Lagrangian mesh approach i.e. a waterbag model \cite{Berk1967} where each contour of constant phase-space density is a waterbag associated with the EPs adiabatic invariants. This enables capturing the particle trapping effect in phase-space and implies that as the separatrix expands and moves due to frequency chirping, the phase-space density of the trapped EPs is set to the ambient distribution at the trapping point.

In section \ref{sec:model}, the model is introduced and the main equations governing the shape of the BGK-type chirping mode and the frequency chirping rate are derived. Simulation data of the BOT code is analysed in section \ref{sec:BOT} from which the initial shape of the coherent phase-space islands is established. The numerical scheme implemented to solve the model equations is briefly given in section \ref{sec:num}. In section \ref{sec:result}, a single-resonance chirping wave with deepening potentials is studied. Therein, specific ranges of fast particles pitch-angles are introduced in which higher particle resonances, in this case $2^{\text{nd}}$, can have a non-negligible contribution to the linear growth rate $(\gamma_l)$ of the wave-particle interaction. Hence, we also report on the impact of higher resonances on the evolution of chirping waves for both up-chirping and down-chirping cases. This is achieved by comparing the evolution of the plane wave potential and the frequency chirping rate for a single and double resonance interaction. We also evaluate the validity of the adiabatic limit for each reported case. Finally, section \ref{sec:sum} is a summary. It is noteworthy that the formalism and the notation presented throughout the manuscript are based on the previously reported model of Ref. \cite{Hezaveh2017} to which the reader is referred for a more detailed derivation.   
\section{Theoretical framework}
\label{sec:model}
We consider the bump-on-tail instability problem of a plasma wave in which the energetic electrons drive the mode marginally unstable until it saturates due to the nonlinear coarse-graining of the electron distribution function in phase-space. Then, if the mode is subject to weak damping into the bulk plasma with a rate denoted by $\gamma_d$, the sideband oscillations are excited and evolve into chirping modes. It is remarkable to mention that fast particles collisions can change the nonlinear evolution of the mode which are neglected here. Hence, the physical picture is a BGK mode with a chirping frequency in a time scale $(t_{\text{slow}})$ much smaller than the bouncing time scale $(t_{\text{fast}})$ of electrons trapped in the mode. Therefore, we have
\begin{equation}
\dv{\omega_b}{t}\ll \omega_b^2,
\label{eq:adb_cnd}
\end{equation}
where $\omega_b$ is the bounce frequency of the electrons trapped in the wave. We consider $\gamma_l \ll \omega_{\text{pe}}$ which implies the separatrix width is much smaller than the characteristic width of the phase-space density and the near-threshold unstable resonance is isolated i.e. overlap of resonances leading to diffusive transport and wave-wave coupling are ignored.

The equilibrium picture of fast electrons dynamics is built by applying a nonuniform static magnetic field. Fast electrons bounce or transit along the field lines. This resembles trapped and passing particles along the field lines in 3D geometries, with the effect of drift orbit width and toroidal precession ignored. The Hamiltonian governing the equilibrium guiding-centre motion of electrons, denoted by $H_{\text{eq}}$, can be derived by applying the Legendre transformation to the gyro-averaged Littlejohn Lagrangian \cite{littlejohn}. This gives
\begin{equation}
H_{\text{eq},\alpha} = \frac{p_{z}^2}{2m_e}-\mu B_0 \cos(k_{\text{eq}}z)+\mu B_c,
\label{eq:Heq}
\end{equation}
\begin{figure}[b!]
  \centering
   \includegraphics[scale=0.55]{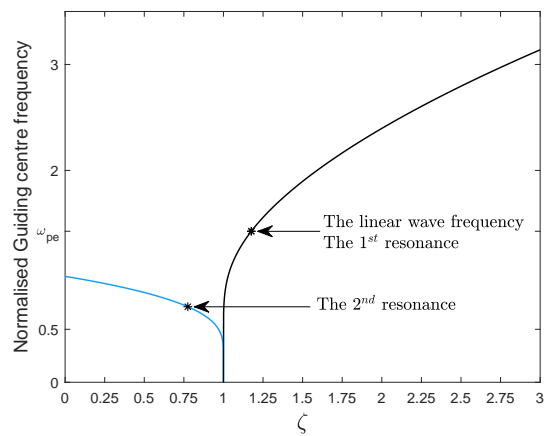}
  \caption{Guiding centre frequency vs. the energy parameter for the fast electrons equilibrium motion.}
  \label{fig:eqfreq}
\end{figure}
where $\alpha$ is a label that denotes the orbit type of the fast electrons motion in the magnetic field based on their pitch angle: throughout the paper, $\alpha=\bf{T}$ and $\alpha=\bf{P}$ represent the trapped and passing electrons in the equilibrium field, respectively, $p_{z}$ is the momentum of energetic electrons aligned with the field, $m_e$ is the electron mass, $\mu$ is the magnetic moment and $k_{\text{eq}}$ denotes the spatial periodicity of the field. The constants $B_0$ and $B_c$ are chosen such that the wave frequency $\omega_{\text{pe}}$ is low compared to the ion cyclotron oscillations and its wavelength is large compared to the electron Larmor radius. Also, it is assumed that all the particles have a single value of the magnetic moment $\mu$. The 1D equilibrium Hamiltonian given by \eqref{eq:Heq} resembles that of a large aspect ratio tokamak i.e. $\epsilon=\frac{r}{R_0}\ll1$, where $\epsilon=\frac{B_0}{B_c}$ is the inverse aspect ratio with $r$ and $R_0$ being the minor and major radius, respectively, and higher order terms in the expansion of the magnetic field in $\epsilon \cos\left (kz\right )$ are neglected. A canonical transformation to action-angle variables ($J_{\alpha},\theta_{\alpha}$) enables a description of the unperturbed motion using $H_{\text{eq},\alpha}(J_{\alpha})$ independent of the corresponding coordinate ($\theta$) which scales linearly with time. Using $\dot{\theta}=\pdv{H_{\text{eq},\alpha}}{J_{\alpha}}$, one can investigate the equilibrium bounce or transit frequency of the fast electrons motion depicted in \fref{fig:eqfreq}. The energy parameter,
\begin{equation}
\zeta = \frac{E+\mu(B_0-B_c)}{2 \mu B_0}
\label{eq:zeta}
\end{equation}
with $E$ being the equilibrium energy, specifies the orbit type of each fast electron. 

In this model, it is assumed that the bulk plasma responds linearly to the field $(U)$ and therefore the corresponding response is found by implementing a perturbative approach to the fluid description. In the presence of the perturbations, the total Hamiltonian of the fast electrons reads
\begin{equation}
H_{\text{total,}\alpha} = H_{\text{eq},\alpha} + U.
\label{eq:H_total}
\end{equation}
In principle, one should implement the Liouville's theorem or the Vlasov equation $\{f,H\}=0$ and either follow the fast electrons trajectories corresponding to the above Hamiltonian i.e. a Lagrangian point approach, or apply a fixed grid discretisation to the phase-space i.e. an Eulerian approach, in order to find the perturbed phase-space density of energetic electrons. Nevertheless, we focus on two separate stages of the wave evolution, namely the linear stage and the nonlinear long range chirping stage. In the former, we resolve the perturbed phase-space density of fast electrons using a linear perturbative analysis while the latter benefits from the Liouville theorem and the adiabatic ordering which enables a Lagrangian contour approach in fast electrons phase-space and construct a non-perturbative approach to find the perturbed density of fast electrons. Here, $f$ is the total distribution function of fast electrons given by $f_{\alpha}=F_{\text{eq},\alpha}+\tilde f_{\alpha}$,  with $F_{\text{eq},\alpha}$ and $\tilde f_{\alpha}$ being the initial and the perturbed parts, respectively. For simplicity, we consider $F_{\text{eq},\alpha}$ to be linear in the energy parameter i.e. $F_{\text{eq},\alpha}=c\zeta_{\alpha}$, where $c$ is a constant. 

We firstly analyse the linear evolution of the plasma wave. This is achieved by finding analytic expressions for  the linear response of the bulk plasma in a single-fluid model and of the fast electrons using the linearised Vlasov equation in a kinetic description. Then, the total Hamiltonian governing the fast electrons dynamics during the adiabatic chirping is described. Hence, we implement a kinetic description for the energetic electrons in the framework of the adiabatic theory and find the corresponding nonlinear contribution. Subsequently, the Poisson equation is fed with the perturbed density of both the fluid and the fast electrons to solve for the nonlinear field of a sideband of the plasma wave during the frequency chirping. At each frequency, the wave potential is a long-term nonlinear solution of the Vlasov-Poisson system, hence a BGK-type wave.

\subsection{Linear evolution of the plasma wave}
For a linear analysis, a perturbative approach is used to find the perturbed density of the bulk and the energetic electrons. Therefore, we represent the wave potential energy $(U)$ and the perturbed distribution function $(\tilde{f}_{\alpha})$ as 
\begin{eqnarray}
U = \sum_{n=1}^{\infty} \frac{e\phi_{n}}{2} \exp \left [ in \left ( k_{p} z-\omega t \right ) \right ] + c.c   \nonumber \\
=  \sum_{n=1}^{\infty}  \sum_{p=-\infty}^{\infty} \frac{e\phi_{n}}{2}V_{\alpha,n,p} \left ( J_{\alpha} \right ) \exp \left [ i \left ( p\theta-n\omega t \right ) \right ] + c.c 
\label{eq:U_pert}
\end{eqnarray}
and
\begin{eqnarray}
\tilde{f}_{\alpha}=\sum_{n=1}^{\infty} \sum_{p=-\infty}^{\infty}\hat{f}_{\alpha,n,p} \left ( J_{\alpha} \right ) \exp \left [i \left  ( p \theta - n \omega t \right ) \right ] + c.c,
\label{eq:f_pert}
\end{eqnarray}
where we have expanded $\e^ { in \left ( k_{p} z-\omega t \right )} $ in action-angle variables of the unperturbed motion i.e.
\begin{equation}
\exp \left [ in \left ( k_{p} z-\omega t \right ) \right ] = \sum_{p=-\infty}^{\infty} V_{n,p}(J) \exp \left [i \left  ( p \theta - n \omega t \right ) \right ],
\label{eq:exp}
\end{equation} 
$\omega=\omega_{r}+i \gamma_{l}$ is the complex frequency, $k_{p}$ the wave-number of the plasma mode, $V_{\alpha,n,p} \left ( J_{\alpha} \right )$ is the orbit averaged mode amplitude which specifies the coupling strength and plays the same role as the matrix elements introduced in Ref. \cite{berk1995,breizman1997}. 
\subsubsection{The bulk plasma response - MHD}
\label{subsec:mhd}
For an isotropic distribution and a uniform density of the bulk plasma along the equilibrium field, we focus on the perturbations along the field lines in which case the equilibrium field does not interact with the bulk plasma. The equation of motion and the linearised continuity equation read
\begin{align}
&\frac{\partial V_c}{\partial t}=-\frac{1}{m_e}\frac{\partial U}{\partial z}- \nu V_c, \label{eq:systemc} \refstepcounter{equation} \subeqn \\
&\frac{\partial \delta n_c}{\partial t}=-n_c \frac{\partial V_c}{\partial z},  \label{eq:systemd} \subeqn
\end{align}
where $U$ is the energy of the electrostatic mode, $\epsilon_{0}$ is the permittivity of free space, $\nu=2\gamma_d$ is the Krook collision frequency of the cold electrons, $V_c$ is the flow velocity of the cold electrons and $n_c$ and $\delta n_c$ are the unperturbed and perturbed density of the cold electrons, respectively. For a linear response, we consider n=1 and substitute Eq. \eqref{eq:U_pert} into Eq. \eqref{eq:systemc} to find $V_c$. Next, Eq. \eqref{eq:systemd} can be implemented to find
\begin{align}
V_{c}=\frac{k_{p}U}{\omega m_{e}}, \refstepcounter{equation}	\label{eq:lfa}	\subeqn \\
\delta n_c = \frac{k_{p}^{2} n_{c} U} {m_{e} \omega^{2}}.  \label{eq:lfb}   \subeqn
\end{align}

\subsubsection{Energetic electrons response - Kinetic description}
\label{subsec:eqmotion}

To first order in perturbations ($n=1$), the fast electron population responds linearly and one can find an analytic perturbative solution,
\begin{equation}
\hat{f}_{\alpha,n=1,p} = \frac{pe \phi_{n=1} V_{\alpha,n=1,p} \left (J_{\alpha}\right) \pdv{F_{\text{eq}} \left (J_\alpha \right )}{J_\alpha}}{2 \left (p\Omega_{\alpha} - \omega \right )},
\label{eq:fpert}
\end{equation} 
to the linearised Vlasov equation, from which one can find the resonance condition $\omega_{r}=p\Omega_{\alpha}$, which if satisfied, fast electrons can resonate with the mode.  Provided that the mode has a non-zero component of the electric field  aligned with the particles guiding centre trajectories, electrons will exchange energy with the mode. Here, $\omega_{r} \approx \omega_{\text{pe}}$ and $p$ is an integer denoting the resonance number. More precisely, $p$ is the number of the Fourier coefficient as a result of expanding the wave equation \eqref{eq:BGKenergy} in AA variable ($\theta_{\alpha}$) of the equilibrium motion. 

The linear perturbative responses of both the bulk plasma and the energetic electrons, represented in Eqs. \eqref{eq:lfb} and \eqref{eq:fpert}, can be substituted in the Poisson equation, given by 
\begin{equation}
\frac{\epsilon_0}{e}\frac{\partial^2 U}{\partial z^2}=-e\left[\sum_{\alpha} \int \tilde{f}_{\alpha} dv +\delta n_c \right], \label{eq:poisson} 
\end{equation}
to find the linear dispersion relation and subsequently the linear growth rate of the initial plasma mode as
\begin{align}
\gamma_{l}=& \frac{\omega_{\text{pe}} \pi e^{2}}{2\epsilon_{0}k_{p}m_{e}} \sum_{\alpha}\sum_{p} \left [ \pdv{F_{\text{eq},\alpha}}{\zeta_{\alpha}} V_{\alpha,n=1,p}^2  \right. \nonumber \\
& \left. \times \abs{\dv{\Omega_{\alpha}}{\zeta_{\alpha}}}_{\Omega_{\alpha} \left ( J_{\alpha} \right ) = \frac{\omega_{\text{pe}}}{p} }^{-1} \right ].
\label{eq:lineargrowthrate}
\end{align}
In the next part, we find the perturbed density of fast electrons during the evolution of the chirping wave and construct the nonlinear equation of the wave potential amplitude. 

\subsection{Chirping waves}
\label{subsec:chirpingwave}

For a dispersion relation of the form $\omega=\omega_{\text{pe}}$ and in a non-perturbative approach subject to the adiabatic limit where the mode evolves slowly, we represent the BGK-type mode with a chirping frequency by
\begin{equation}
U[z,t]=\sum_{n} A_{n}(t) \cos \left [n \left (k_{p} z-\phi\left (t\right ) \right ) \right ],
\label{eq:BGKenergy}
\end{equation}
where the wave oscillates on a fast time scale on the order of $\omega_{\text{pe}}^{-1}$ whereas its envelope $A_n$, as the Fourier coefficient of the $\text{n}$-th harmonic, evolves on a slow time scale subject to the adiabatic ordering
\begin{equation}
\dv{\ln A_n}{t}\ll\dot{\phi}\left (t_{\text{fast}}\right ).
\label{eq:adbforAn}
\end{equation}
It is noteworthy that for dispersion relations of the form $\omega= c k_{p}$, where $c$ is a constant, Eq. \eqref{eq:BGKenergy} represents a sum over linear modes and subsequently alternative discretisation methods should be used. 

The nonlinear dynamics of the fast electrons can be described in a frame co-moving with the wave and this leaves us with a time-dependent Hamiltonian that evolves adiabatically in time. Using \eqref{eq:H_total} and \eqref{eq:BGKenergy}, this time-dependent Hamiltonian is written as
\begin{eqnarray}
H_{\text{total,}\alpha} =& \frac{1}{2} \left. \pdv[2]{H_{0,\alpha}}{\tilde{J}_{\alpha}}  \right |_{\tilde{J}_{\alpha} = \tilde{J}_{\text{res},\alpha} \left (t \right )}  \left (\tilde{J}_{\alpha}-\tilde{J}_{\text{res},\alpha}   \left (t\right ) \right )^2 +  \nonumber \\
&\frac{1}{2} \sum_{n} A_{n}\left (t \right ) V_{\alpha,n,n} \exp \left (in\tilde{\theta} \right ) + c.c,
\label{eq:newhamiltonianb}
\end{eqnarray}
where a canonical transformation as
\begin{align}
&\tilde{\theta}_{l}=l\theta-\phi \left (t\right ) , \label{eq:ct} \refstepcounter{equation} \subeqn \\
&\tilde{J}_{\alpha} = \frac{J_{\alpha}}{l},  \label{eq:ct} \subeqn
\end{align}
is implemented to transfer the coordinates to a frame co-moving with the wave and cancel the fast time dependency, the wave potential energy ($U$) of Eq. \eqref{eq:BGKenergy} has been Fourier decomposed in AA variables of the unperturbed motion with $V_{\alpha,n,p}$ denoting the Fourier coefficients, $p$ is a label that denotes the resonance number for the linear perturbations (n=1) whereas in the nonlinear case, $l=\frac{p}{n}$ identifies the resonance number. The above Hamiltonian is expanded around the middle of the chirping wave trapping region (separatrix) specified by $\tilde{J}_{\text{res},\alpha}$ and assumes infinitesimal detuning for the energetic electrons bouncing in the trapping region of the wave, $V \approx V(\tilde{J}_{\text{res}})$.

For such a system, the lowest order term corresponding to the expansion of the adiabatic invariant in the small parameter $\beta$, as the proportion of the bounce period of the electrons trapped in the chirping wave to the slow time scale of the mode evolution), is commonly taken to be the action \cite{cary1986,cary1989}, which reads
\begin{equation}
I = \frac{1}{2\pi}\int \tilde{J} d\tilde{\theta}, 
\end{equation}
\begin{figure}[b!]
  \includegraphics[scale=0.58]{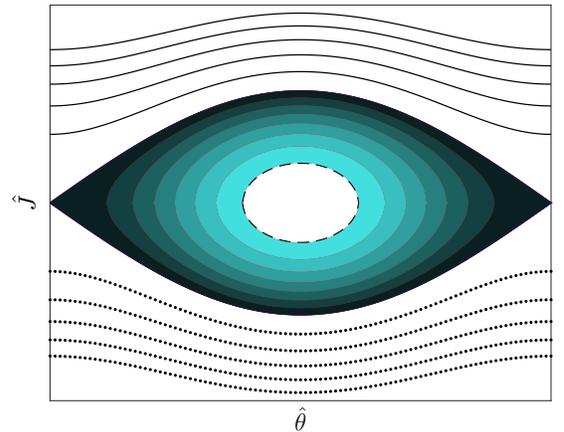}
  \caption{An expanded phase-space island. The unshaded area inside the separatrix represents the initial island just after the explosive formation stage. The dashed area illustrates the phase-space waterbags as contours of the distribution function.}
\label{fig:exp_island}
\end{figure}
where the integration is performed from $0$ to $2\pi$ over the angle variable. Conservation of the adiabatic invariants implies that the corresponding phase-space area occupied by each adiabatic invariant is conserved. This means that in a discretised picture, the  phase-space area between adiabatic invariants, denoted by $A_i$ is preserved as the wave chirps. \Fref{fig:exp_island} shows the phase-space of an expanded separatrix. The unshaded area surrounded by the dashed curve in the middle of the separatrix corresponds to the initial separatrix i.e. the shape of phase-space structures just after formation and prior to the adiabatic evolution. Each shaded region $(A_i)$ is the area between two adjacent adiabatic invariants $(I_i,I_{i+1})$. In addition, in the absence of collisions, the number of electrons $(N_i)$ in the area $A_i$ remains fixed during the frequency chirping. The integral form of the Liouville theorem reads
\begin{equation}
\int_{I_i}^{I_{i+1}} f_idA_i = N_i = \int_{I_i}^{I_{i+1}} f_i^{\prime}dA_i^{\prime}
\end{equation}
where $f_i$ is the distribution function of electrons in $A_i$ and the primes denote the values after the motion of an island in phase-space during frequency chirping. Under the adiabatic ordering and taking an infinitesimal width for $A_i$ by choosing small time steps, the fast bounce frequency of the trapped electrons in the BGK mode allows one to assume $f_i$ to be the same across $A_i$, which gives
\begin{equation}
f_i \times A_i = N_i = f_i^{\prime} \times A_i^{\prime}.  
\end{equation}
The preservation of the adiabatic invariants explained above ensures $A_i=A_i^{\prime}$ which guarantees that the distribution function remains constant in between adjacent adiabatic invariants i.e a phase-space waterbag. This implies that instead of taking an Eulerian grid approach of solving the Vlasov equation or a Lagrangian approach to solve the equations of motion for each particle to resolve the perturbed phase-space density, we can define a set of Lagrangian contours along which the density remains fixed.

The trapped electrons in the BGK mode move with the wave in phase-space as the frequency chirps and hence have the dominant contribution to the perturbed density $(\tilde{f})$ which equals the difference between the value of distribution function at that point and the ambient distribution. Therefore, we can find $\tilde{f} =f_0 - F_{\text{eq}}(t) = F_{\text{eq}}(t=0) - F_{\text{eq}}(t)$ for each point inside the separatrix and $\tilde{f}=0$ otherwise. Here, $f_0$ is the lowest order term of the expansion of $f$ around $\beta$. 
Similarly, one can bounce-average the Vlasov equation under the adiabatic ordering to derive the above expression (see \cite{Nyqvist2013,Hezaveh2017}). Consequently, we discretise the phase-space area inside the wave trapping region using the adiabatic invariants of the fast electrons and hence create level sets of the distribution function in phase-space i.e. a stepped distribution profile. Now, the problem of resolving the perturbations in fast electrons population during frequency chirping is 
framed as tracking the dynamics of the phase-space curves corresponding to the adiabatic invariants.
 
At this stage, the expressions \eqref{eq:lfb} and \eqref{eq:BGKenergy} can be substituted into the Poisson equation to solve for the field 
\begin{eqnarray}
A_{n} & \left(t\right)= \frac{1}{2\pi k_{p} n_{c}} \left [ \frac{\omega^{2}}{n^{2}\hat{\omega}^{2}-1}\right ] \sum_{\alpha} \sum_{l} \int_{0}^{2\pi} \int_{0}^{\infty} \left [ \tilde{f}_{\alpha,l} (\tilde{\theta}_l,\tilde{J}_l ) \right. \nonumber \\
&\left. \times  V_{\alpha,n,p \cross l}\exp ( in\tilde{\theta}_l) + c.c \right ]  d\tilde{J}_l d\tilde{\theta}_l.
\label{eq:Fcoef1} 
\end{eqnarray}
The above expression sets the nonlinear integral equation for each Fourier coefficient, $A_{n}(t)$, which takes into account the contribution of fast electrons with different orbit types as well as the higher resonances, denoted by the sum over $l$. It is noteworthy that the summation over different resonances is removed in Ref. \cite{Hezaveh2017}.

The chirping mechanism is based on extracting energy from the fast particles distribution and deposit it into the bulk plasma. Equating the energy released by the phase-space structure(s) with the energy deposited into the bulk gives
\begin{equation}
\dv{\omega \left (t\right )}{t} =-\left [ \frac{\nu n_{c} \pi k_{p}}{\omega^{3} m_{e}} \sum_{n} A_{n}^2 \left(t\right ) \right] \frac{1}{\sum_{\alpha,l} N_{\alpha,l} \left ( \dv{\Omega_{\alpha,l}}{J_{\alpha,l}} \right )^{-1}}.
\label{eq:srate} 
\end{equation}
The total number of the particles inside each coherent structure reads
\begin{eqnarray}
N_{\alpha,l} = \frac{1}{m_e} \int_{0}^{2\pi}\int_{\tilde{J}_{\alpha,max-}}^{\tilde{J}_{\alpha,max+}} \tilde{f}_{\alpha,l} \left (\tilde{J}_{\alpha,l},\tilde{\theta}_l \right ) d\tilde{J}_{\alpha} d{\theta}.
\label{eq:N}
\end{eqnarray}
In general, $\tilde{f}$ depends on the phase-space coordinates $(\tilde{J},\tilde{\theta})$ and a numerical treatment of the phase-space integral is required. For a growing separatrix, the newly trapped electrons inside the separatrix, specified by their adiabatic invariants, will carry the ambient phase-space density at the time of trapping. Therefore, in a time-discretised scheme, such a phase-space structure consists of an initial shape, which corresponds to the time when holes/clumps are just formed, surrounded by Lagrangian contours (waterbags) having different phase-space densities (see \fref{fig:exp_island}). As the separatrix expands, phase-space waterbags with uniform distribution functions are added around the initial separatrix.    
\begin{figure*}[b!]
\centering
\subfloat[]{
\includegraphics[scale=0.57]{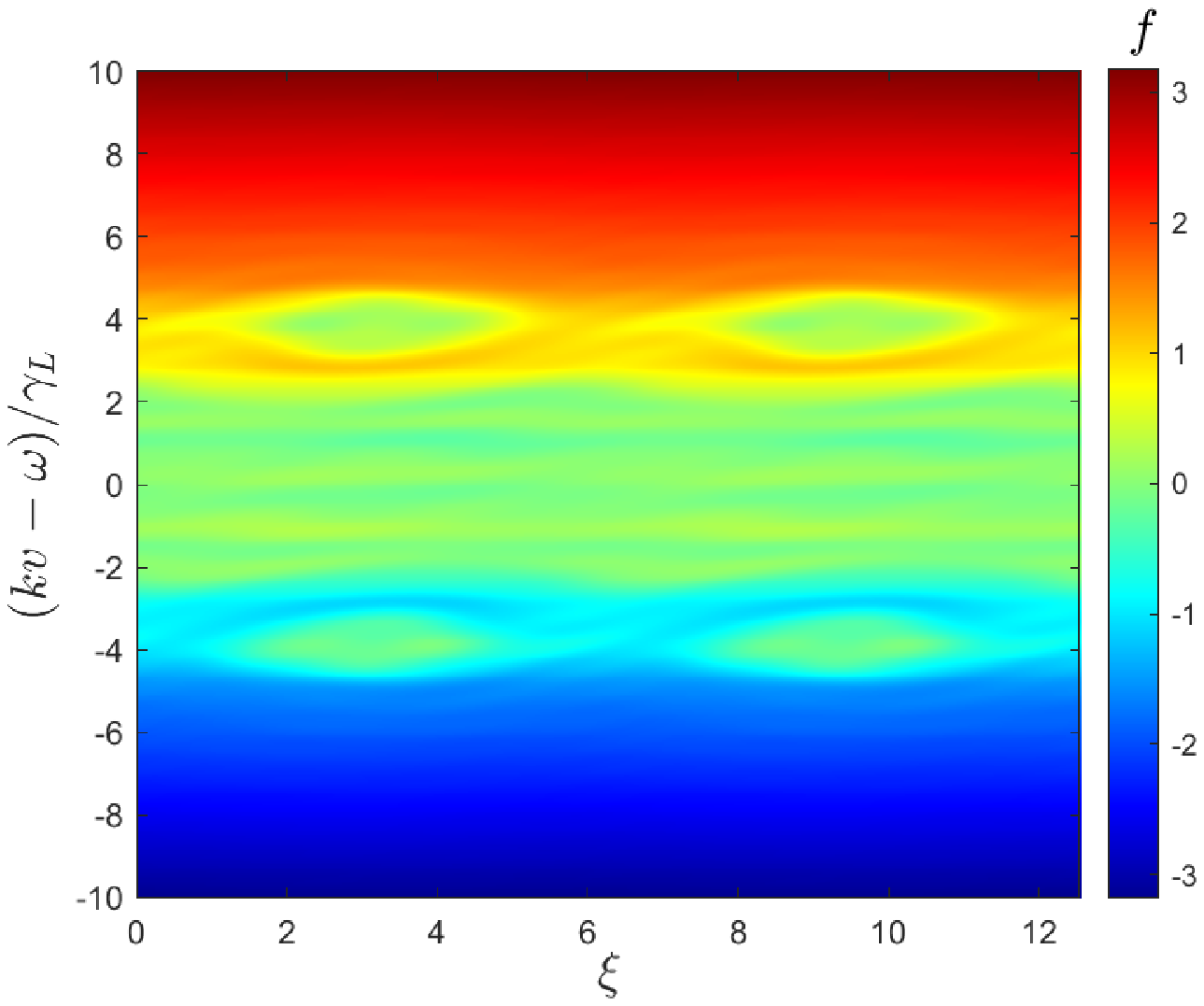}
\label{fig:ps}
}
\subfloat[]{
\includegraphics[scale=0.57]{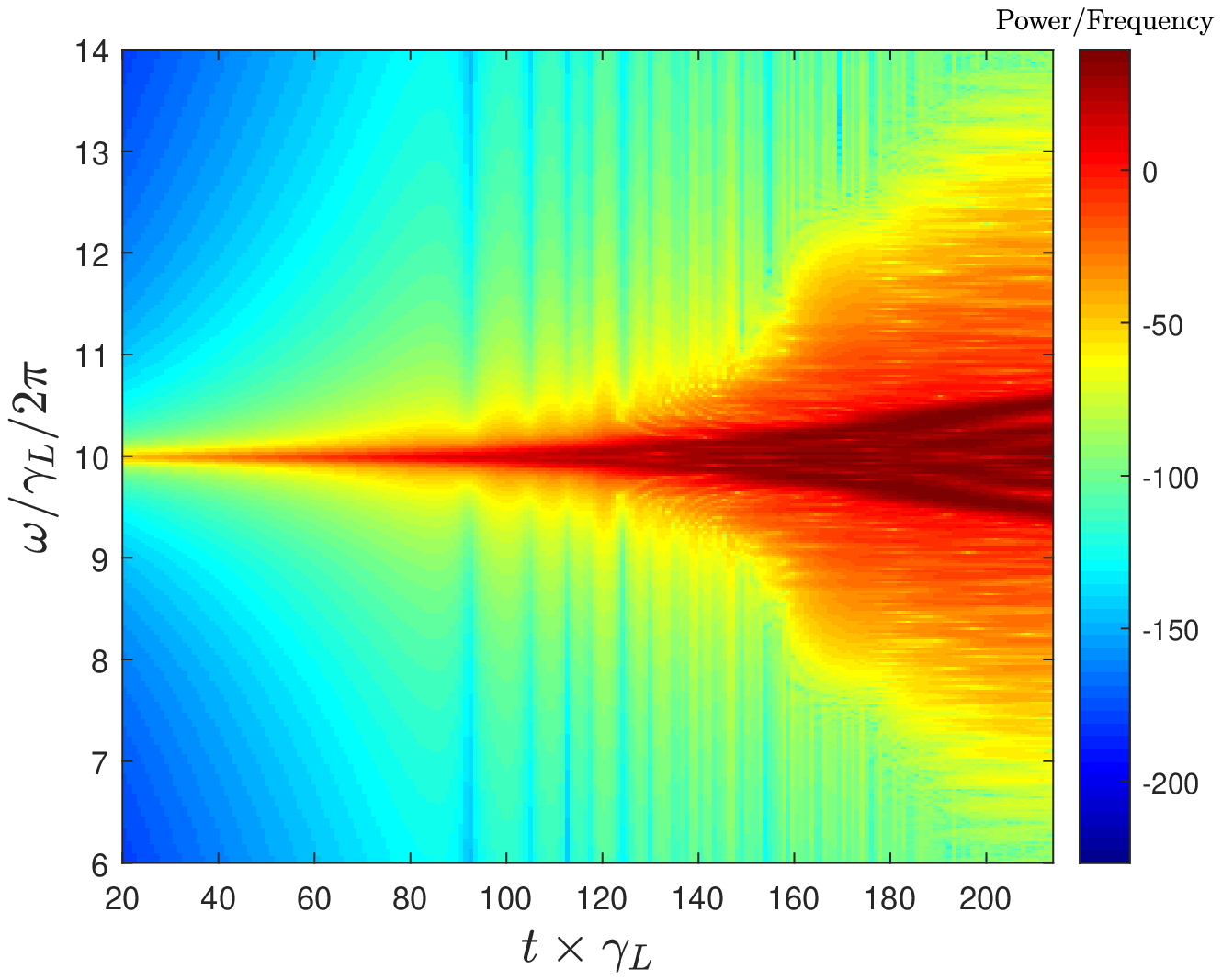}
\label{fig:spc}
}
\caption{Formation of islands in the phase-space of energetic particles (a). The short-time Fourier transform of the wave signal (b). Panel a corresponds to the last time slice of panel b where the frequency has chirped to $\approx 5.5\%$ of its initial value.}
\label{fig:ps_spc}
\end{figure*}

So far, we have set the necessary tools to investigate the evolution of the chirping wave and a numerical approach is required to solve \eqref{eq:srate} along with \eqref{eq:Fcoef1}. However, as shown in Refs. \cite{Boris2010,Hezaveh2017} and discussed in Ref.\cite{Nyqvist2013}, evaluation of \eqref{eq:srate} at early stages of frequency chirping in this model reveals a square root dependency of the frequency on time. This dependency implies that the adiabatic condition is never formally satisfied for very early stages of chirping. In addition, for an expanding phase-space island, this may result in numerical errors due to large particle trapping at the early stages. In order to tackle this issue, we use the following facts:
\begin{itemize}
  \item The holes and clumps form off the initial resonance \cite{Lilley2014},
  \item The violation of the adiabatic condition occurs over a very short period and this is implied by the condition $\gamma_l \ll \omega_{\text{pe}}$,
  \item The adiabatic condition will remain valid once its satisfied \cite{Hezaveh2017}.
\end{itemize}
These enable solving the system somewhat off resonance by considering an initial shift to the eigenfrequency. This frames the question of what shape the phase-space island will take after the initial shift. In other words, subsequent to an imposed frequency shift to the linear resonance, an appropriate description of the phase-space density is required for the unshaded phase-space area encircled by the dashed curve depicted in \fref{fig:exp_island}. At this point, the challenge concerns the fact that holes/clumps are formed on a characteristic time scale in the order of the bounce period. Thereby, full-scale modelling is required and one can not invoke the adiabatic ordering and Liouville  theorem to avoid following the particle dynamics on the fast time scale i.e. $\omega_b^{-1}$. As a result, we perform simulations using the BOT code to resolve the dynamics during the fast formation stage. This part is covered in the next section where we prescribe an appropriate initialisation for the system.

\section{Implementation of the BOT code for phase-space initialisation}
\label{sec:BOT}

In this section, the procedure taken to find a realistic shape for holes/clumps (phase-space structures) using the simulation data is detailed. The BOT code is an open source Vlasov solver which resolves the evolution of an unstable plasma wave in a bump-on-tail model. It also captures EPs collisions of Krook, drag and diffusion type which has been used to study the effect of dynamical friction force \cite{Lilley2009} and the formation process of holes and clumps \cite{Lilley2014}. In BOT code, the angular dependency ($\cos(kz)$) of the linear plasma wave and its subsequent sidebands oscillations are fixed to be sinusoidal and do not evolve. On the other hand, as a result of the excitation of the sidebands and damping into the bulk plasma, the frequency of the BGK-type chirping modes deviates from the initial eigenfrequency. As this occurs, the nonlinear contribution of the EPs current modifies the sinusoidal mode and adds nonlinearity to the angular shape of the mode. This phenomenon is not captured in the BOT code. 
\begin{figure*}[b!]
\centering
\subfloat[]{
\includegraphics[scale=0.6]{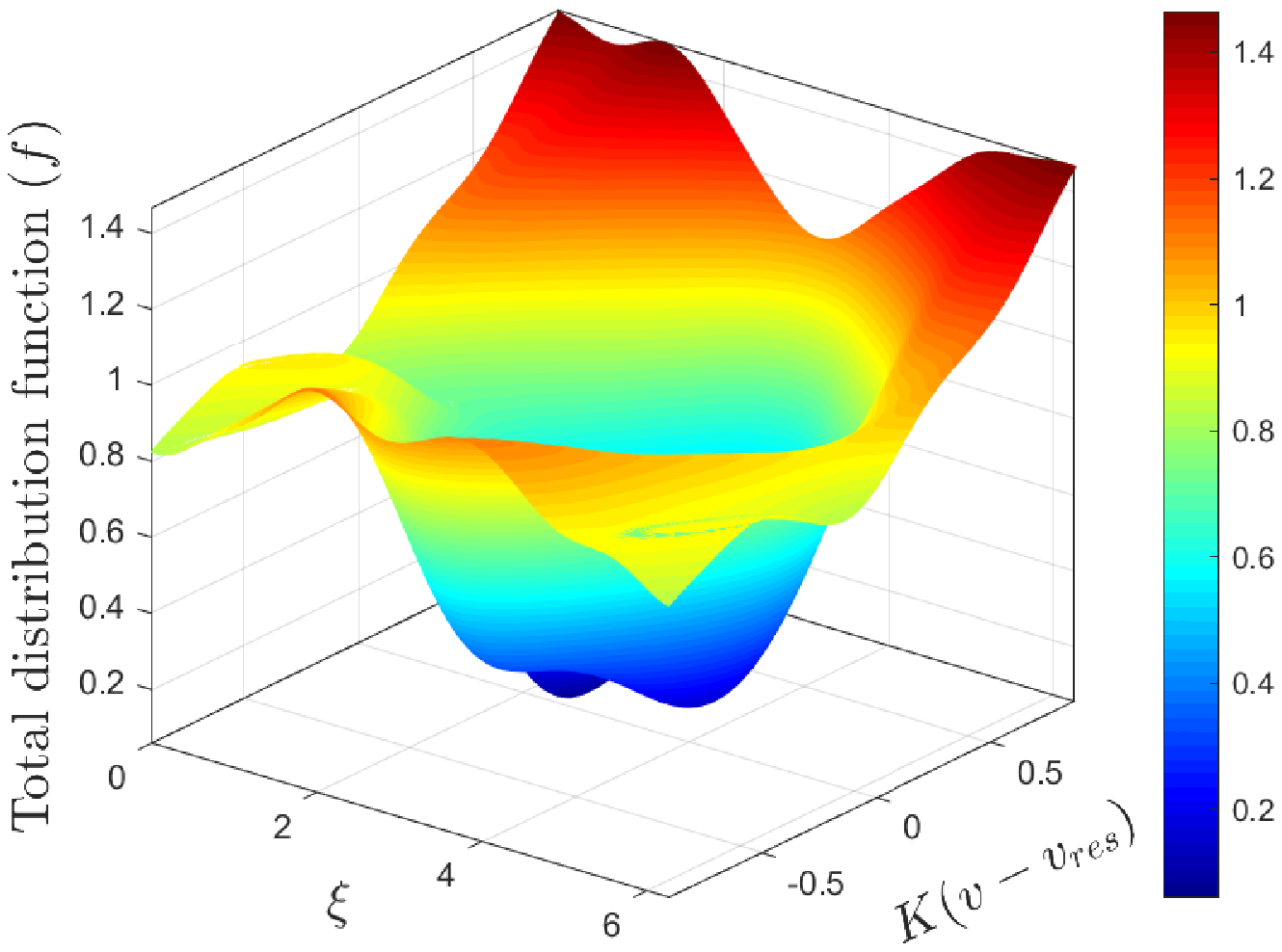}
}
\subfloat[]{
\includegraphics[scale=0.62]{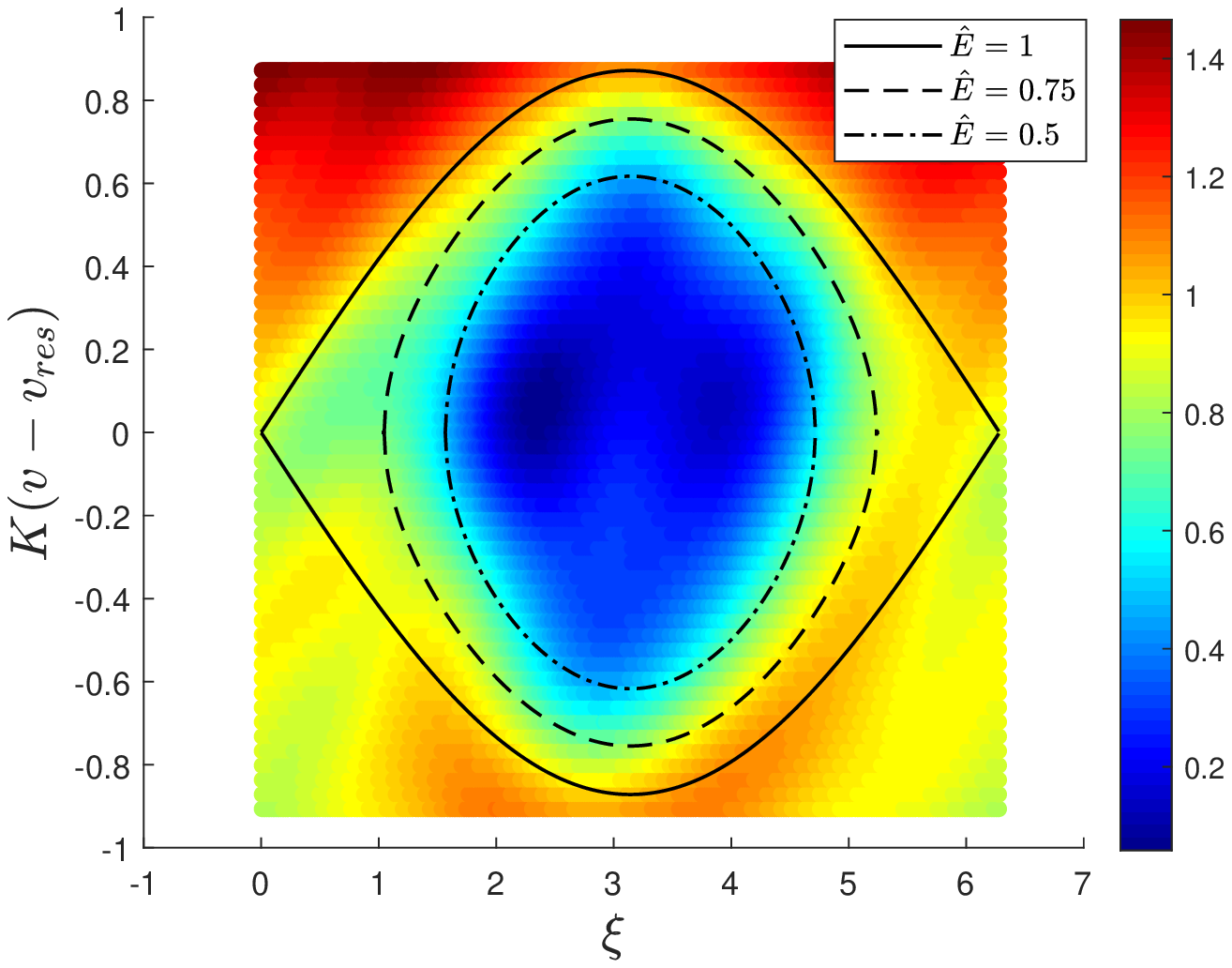}
}
\caption{(a) A Phase-space island (hole) in the BGK wave-frame formed at the top of the flattened region in \fref{fig:ps_spc}. (b) The contour plot of the phase-space density inside the hole shown in panel a. The curves represents contours of constant normalised energy given by \eqref{eq:norm_energy}.}
\label{fig:ps_contour}
\end{figure*}
A consequence of this is the phase-space structures being perfectly eye-shaped. However, for short deviations of the frequency from the initial eigenfrequency the change of the mode shape is negligible (see Ref. \cite{Hezaveh2017}) and the simulation data remains valid for short ranges of frequency chirping. On the other hand, as mentioned in the previous section, holes/clumps form on a time scale comparable to the bounce period and hence the formation process occurs in a short range of frequency sweeping. Consequently, the phase-space analysis of the BOT code at the very early stages of chirping can be used to identify the structure of holes/clumps in phase-space just after their formation and prior to the adiabatic evolution. This information can be implemented to initialise the phase-space of our adiabatic model which can handle long range frequency deviations. In the following, we base our calculations around the time where holes/clumps (phase-space structures) are just formed and the phase-space of the adiabatic model is initialised accordingly. 

The simulation results of the BOT code are illustrated in \fref{fig:ps_spc}, where the phase-space of energetic electrons is demonstrated (\fref{fig:ps}) after the saturation and nonlinear phase-mixing of the electrons when the sideband oscillations has just been excited. The corresponding frequency evolution of the plasma wave is shown in \fref{fig:spc} where it can be observed that the frequency has swept $\approx 5.5\%$ of the initial eigenfrequency $ (\delta \omega_0 = \frac{\Delta \omega_0}{\omega_{\text{pe}}}\approx 5.5\% )$. The phase-space density inside the structures can be used to find an approximated shape for the holes/clumps just after their explosive formation process in order to initialise the adiabatic model for the evolution of these structures. To perform this simulation using the BOT code, the value of $\gamma_d/\gamma_l$ is set to be $0.9$ as a near-threshold instability case and the collisional coefficients are set to zero.  

In \fref{fig:ps_contour}, the structure of an up-chirping hole is depicted together with the contour plot of the phase-space density. As the frequency evolves, snapshots of phase-space reveal that for a fixed wave amplitude, the phase-space density remains the same along the contours of constant energy in the wave frame. However, it is noteworthy that as the frequency evolves, there is a subsequent change in the amplitude of the BGK-type wave and $\tilde{J}_{\text{res}}$. Therefore, the functional dependency between the adiabatic invariant and energy of the trapped electrons in the wave does not remain the same during frequency chirping. However, since the structures are evolving adiabatically, conservation of the adiabatic invariants of the system ensures that the phase-space density remains constant in between the adiabatic invariants. Consequently, we discretise the phase-space using the adiabatic invariants with each region having a constant distribution; a stepped distribution profile (a waterbag model) as a function of the adiabatic invariants for the numerical analysis. 

\begin{figure}[t!]
  \includegraphics[scale=0.58]{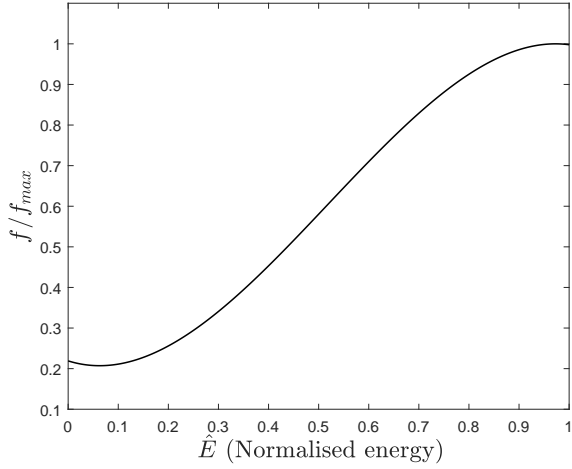}
  \caption{The shape of the phase-space structure found using the BOT code simulations}
\label{fig:shape_hole}
\end{figure}

For each 2D phase-space element of \fref{fig:ps_contour}, the Hamiltonian \eqref{eq:newhamiltonianb} can be utilised to find the corresponding energy in the wave denoted by $E_{\text{total}}$. Subsequently, a polynomial fitting to the data gives the shape of the distribution function inside the phase-space structure which is illustrated in \fref{fig:shape_hole}. The normalised energy $\hat{E}$ is defined as
\begin{equation}
 \hat{E}=  \frac{E_{\text{total}}-U_{\text{min}}}{U_{\text{max}}-U_{\text{min}}},
\label{eq:norm_energy} 
\end{equation}
where $U_{\text{max}}$ and $U_{\text{min}}$ denote the maximum and minimum potential energy of the chirping wave, respectively. This prescribed shape is implemented in section \ref{sec:result} as the initial shape of holes and clumps which start evolving from $\approx 5.5\%$ off the initial resonance. In what follows, the phase-space structures are initialised according to the shape of \fref{fig:shape_hole}.

\section{Numerical Algorithm/procedure}
\label{sec:num} 

In this part, The numerical algorithm implemented to solve the system equations is explained and we introduce the normalisation used on the system equations. For the purpose of normalisation, we firstly need to evaluate \eqref{eq:Fcoef1} and \eqref{eq:srate} at the early stages of frequency chirping. In the limit $(t\approx0)$, the plane wave is still almost sinusoidal/linear $(A_{n\ge2}\approx0)$. Regardless of whether the separatrix is expanding or shrinking and for the case where only magnetically passing electrons $(\alpha=\bf{P})$ contribute to the EPs current through the first resonance $(l=1)$, one can analyse the integral Eq. \eqref{eq:Fouriercoef1} at $t\approx0$ to find
\begin{equation}
A_{1,0} =  \frac{4\omega_{\text{pe}}^{2} \pdv{F_{\text{eq},\bf{P}}}{\zeta_{\bf{P}}} \left. \pdv{\zeta_{\bf{P}}}{\hat{\omega}} \right |_{\hat{\omega}=1} }   {3\pi k_{p} n_{c} } V_{{\bf{P}},1,1,0} \tilde{J}_{\text{max},\bf{P},0} 
\label{eq:initialcoef1} 
\end{equation}
with $\tilde{J}_{\text{max},\bf{P},0}$ being the maximum half width of the saturated/initial trapping region (separatrix at $\tilde{\theta}=\pi$) of the BGK mode corresponding to the first resonance with magnetically passing electrons, given by
\begin{eqnarray}
\tilde{J}_{l,\text{max},\alpha=\bf{P},0} \left (\tilde{\theta}=\pi \right ) = 2\sqrt{\frac{A_{n=1,0}V_{{\bf{P}},1,1,0}}{\abs{\Delta_{\alpha=\bf{P}}}_{t=0}}}.
\label{eq:initalJ} 
\end{eqnarray}
It is worth mentioning that the trapping region of the BGK mode carrying the magnetically trapped electrons has a phase shift of $\pi$ with respect to the one corresponding to the magnetically passing ones (see fig.2 in Ref. \cite{Hezaveh2017}). 
Now, we use $A_{1,0}$ to normalise Eq. \eqref{eq:Fcoef1}. This gives
\begin{eqnarray}
\hat{A}_{n} &\left(t\right)= \frac{3\hat{\omega}^{2}}{8(n^{2}\hat{\omega}^{2}-1)\pdv{F_{\text{eq},\bf{P}}}{\zeta_{\bf{P}}} \left. \pdv{\zeta_{\bf{P}}}{\hat{\omega}} \right |_{\hat{\omega}=1}} \sum_{\alpha} \sum_{l} \int_{0}^{2\pi} \int_{0}^{\infty}   \nonumber \\
&\left [ \tilde{f}_{\alpha,l} (\tilde{\theta}_l,\tilde{J}_l )  \hat{V}_{\alpha,n,p \cross l}\exp ( in\tilde{\theta}_l) + c.c \right ]  d\hat{\tilde{J}}_l d\tilde{\theta}_l,
\label{eq:Fouriercoef1} 
\end{eqnarray}
where $\hat{A}_{n}=\frac{A_{n}}{A_{1,0}}$, $\hat{\omega}=\frac{\omega}{\omega_{\text{pe}}}$, $\hat{V}_{\alpha,n,p \cross l}=\frac{V_{\alpha,n,p \cross l}}{V_{{\bf{P}},1,1,0}}$ and $\hat{\tilde{J}}_l = \frac{\tilde{J}_l}{\tilde{J}_{l,\text{max},\alpha=\bf{P},0}}$.

Using \eqref{eq:lineargrowthrate}, \eqref{eq:initialcoef1} and the normalised time $\tau=\frac{\nu}{3}(\frac{16\gamma_l}{3\pi^2\omega_{\text{pe}}})t$, one can normalise the differential equation \eqref{eq:srate} and find
\begin{eqnarray}
\dv{\left (\hat{\omega} -1\right )^2}{\tau} &= \left [\frac{8\left ( \hat{\omega} -1 \right ) \pdv{F_{\text{eq},\bf{T}}}{\zeta_{\bf{T}}}\left.\pdv{\zeta_{\bf{T}}}{\hat{\omega}}\right |_{\hat{\omega}=1}}{\hat{\omega}^{3}} \sum_{n} \hat{A}_{n}^2 \left(t\right)\right ]  \nonumber \\
&\times \frac{1}{\sum_{\alpha} \sum_{l} \int_{0}^{2\pi}\int_{\hat{\tilde{J}}_{\alpha,max-}}^{\hat{\tilde{J}}_{\alpha,max+}} \tilde{f}_{\alpha,l}  d\hat{\tilde{J}}_{\alpha} d{\theta} \hat{\Gamma}_{\alpha}^{-1}}.
\label{eq:sratespecial} 
\end{eqnarray}

The frequency of the chirping waves is evolved using the above ODE by a $4^{th}$ order Runge-kutta method. At each time step, the nonlinear field is solved by performing iterations on the Fourier coefficients using \eqref{eq:Fouriercoef1}. In each iteration the phase-space integral is resolved numerically by a 2D trapezoidal rule. Energetic electrons are labeled using their adiabatic invariants. Hence each separatrix is identified/discretised using an array of adiabatic invariants and a corresponding array of distribution function values which are initialised using the BOT code data. At the end of each time step, these arrays are updated depending on whether the separatrix is shrunk or expanded.  

\begin{figure*}[h!]
\centering
\subfloat[]{
\includegraphics[scale=0.36]{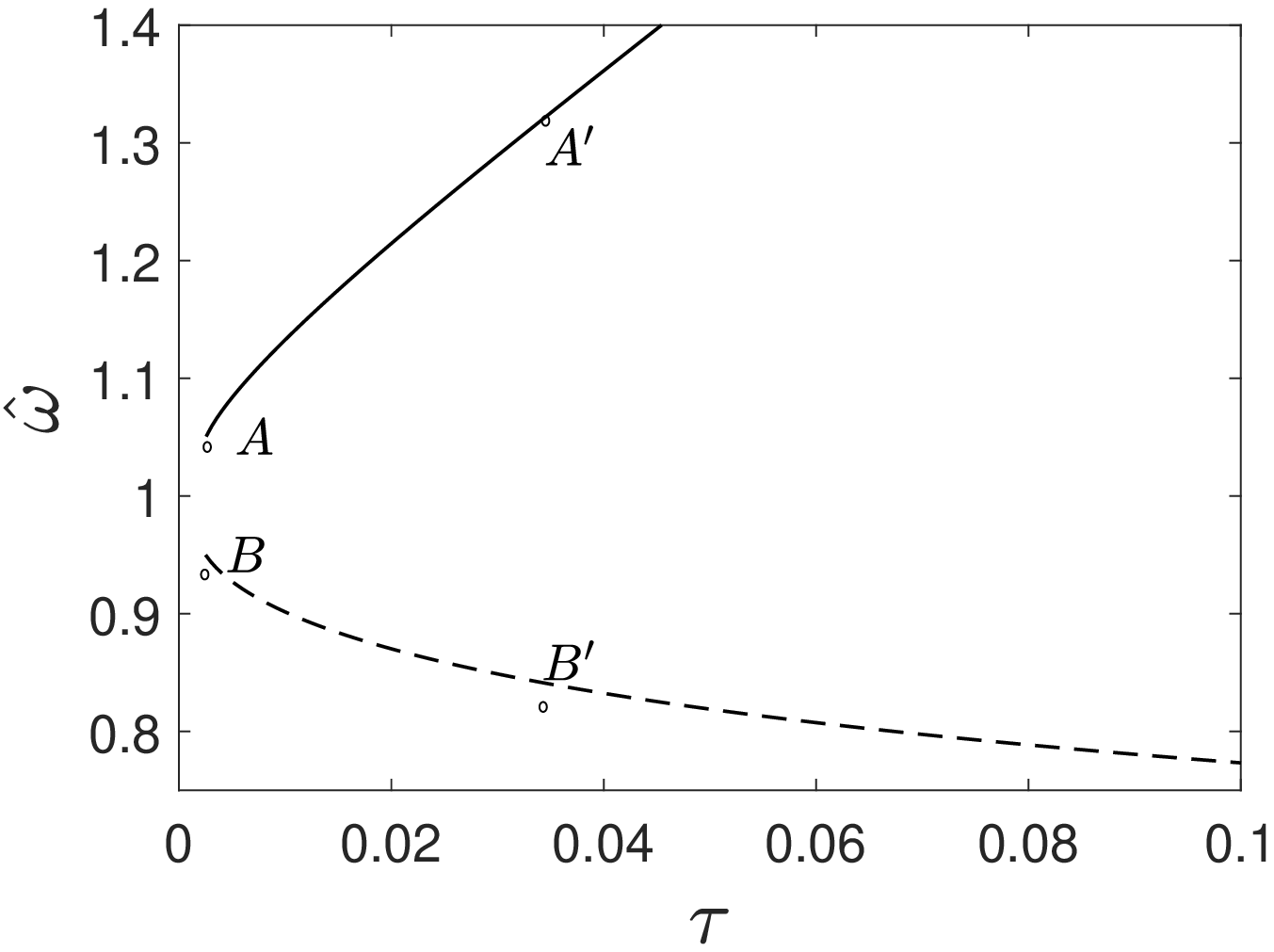}
\label{fig:rate_s}
}
\subfloat[]{
\includegraphics[scale=0.36]{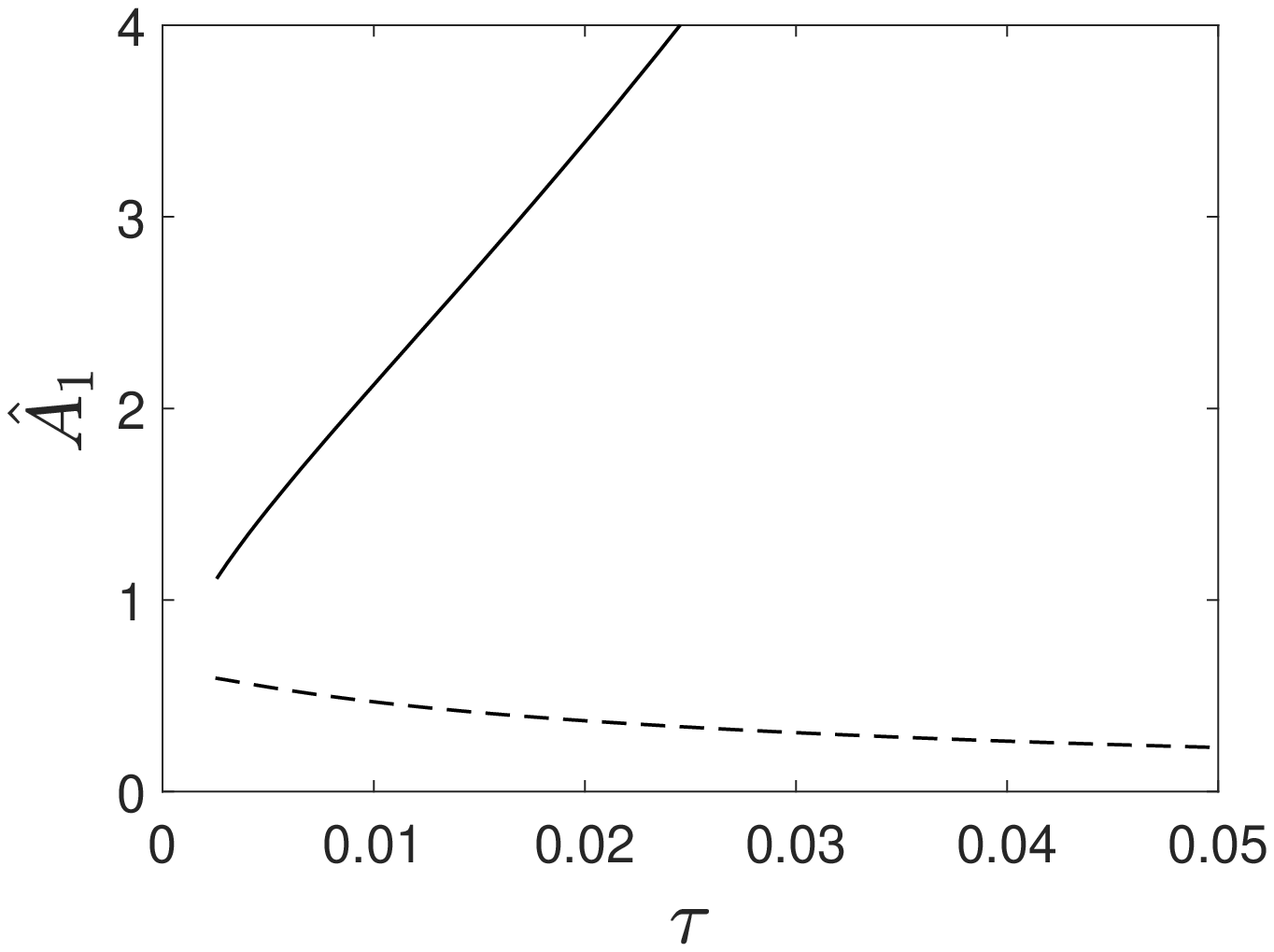}
\label{fig:A_s}
}
\subfloat[]{
\includegraphics[scale=0.36]{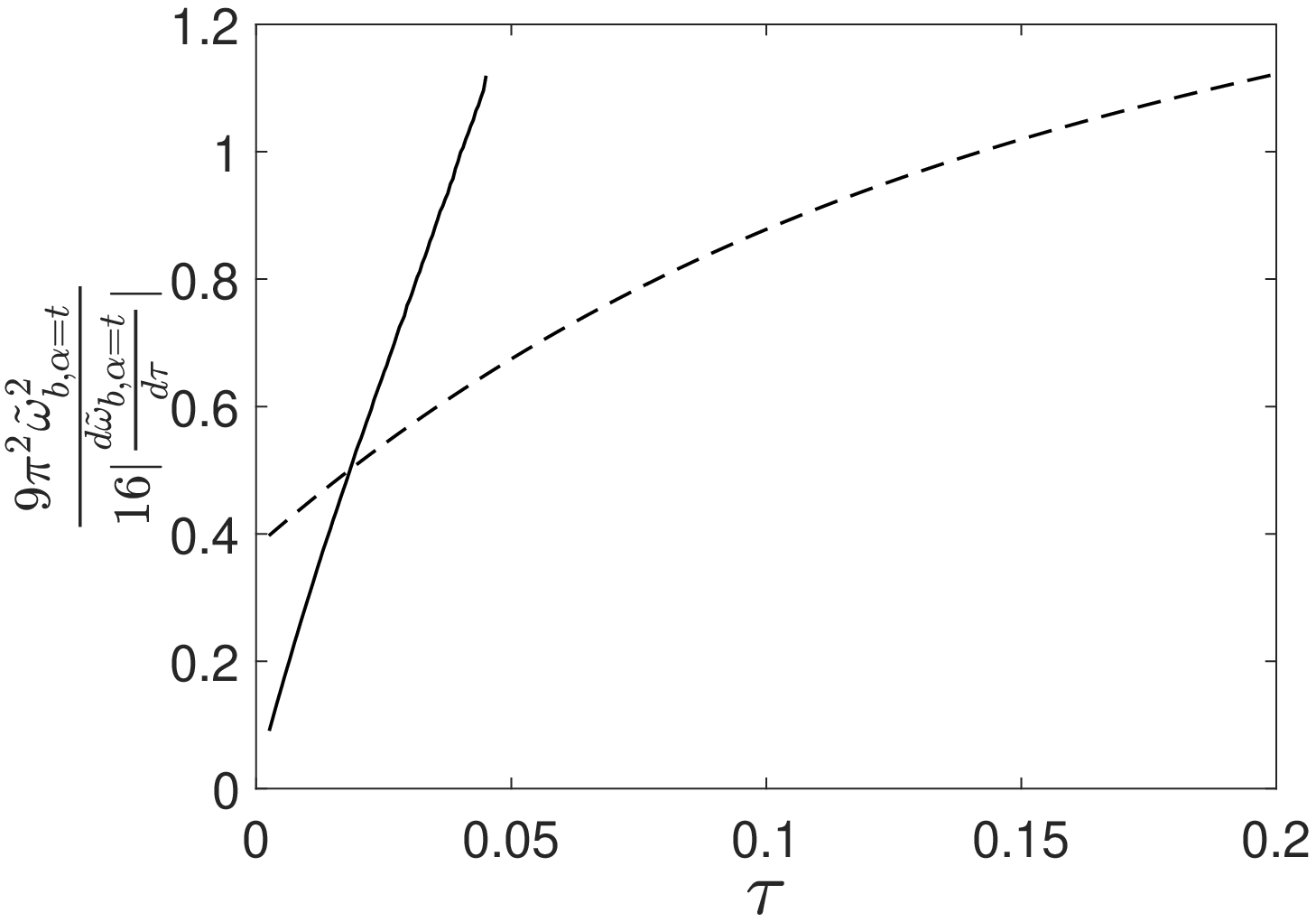}
\label{fig:RHS_s}
}
\caption{Evolution of (a) the frequency, (b) the first Fourier coefficient as a measure of the chirping wave amplitude and (c) the RHS of the adiabatic condition given by \eqref{eq:adb_cnd2} for a single resonance. An initial frequency shift of $\delta \omega_0=0.055\omega_{\text{pe}}$ is considered which corresponds to $\tau_0=0.0025$ following the square root dependency.}
\label{fig:evl_SR}
\end{figure*}
\begin{figure}[b!]
\centering
\subfloat[]{
\includegraphics[scale=0.27]{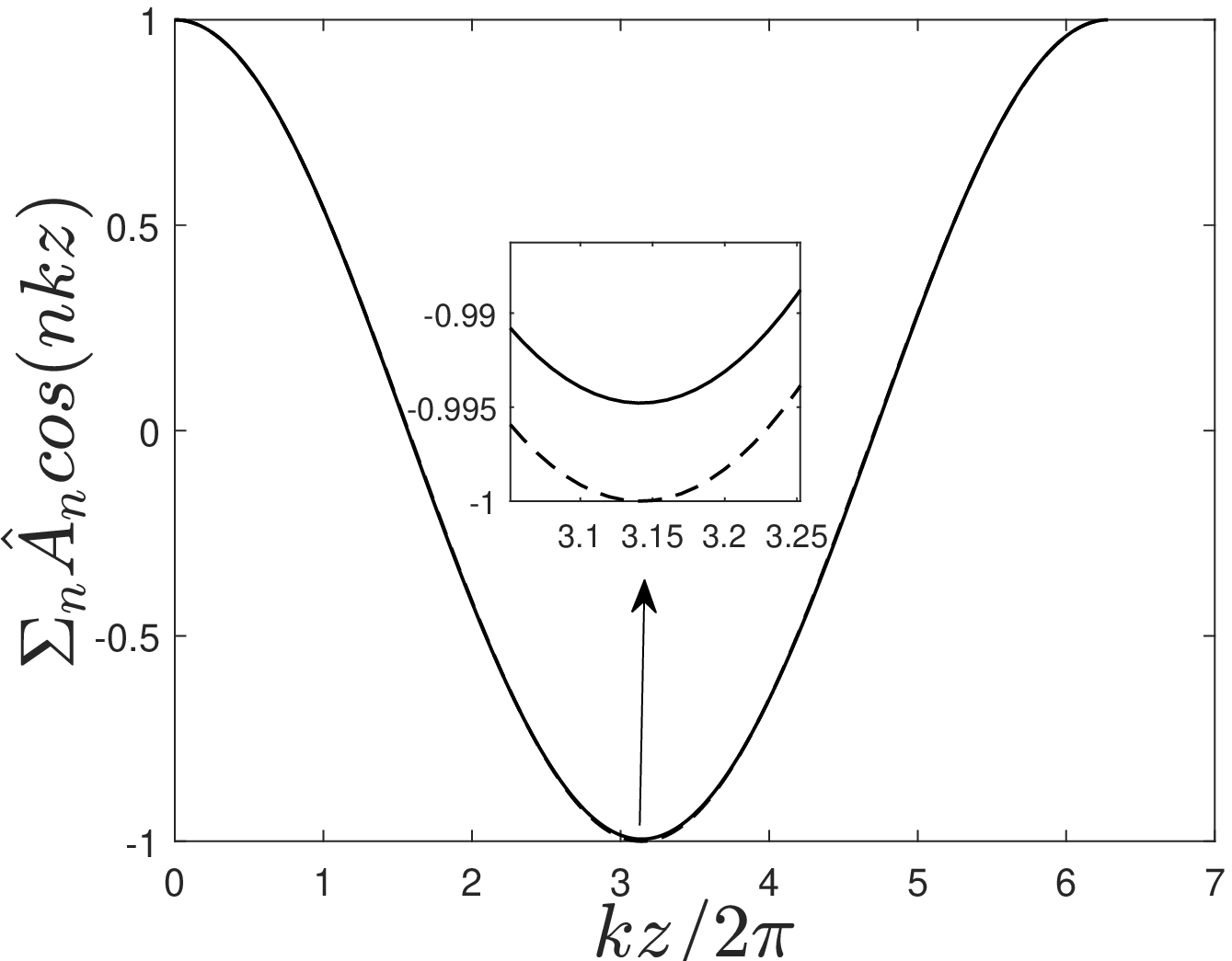}
\label{fig:str_sng_u}
}
\subfloat[]{
\includegraphics[scale=0.27]{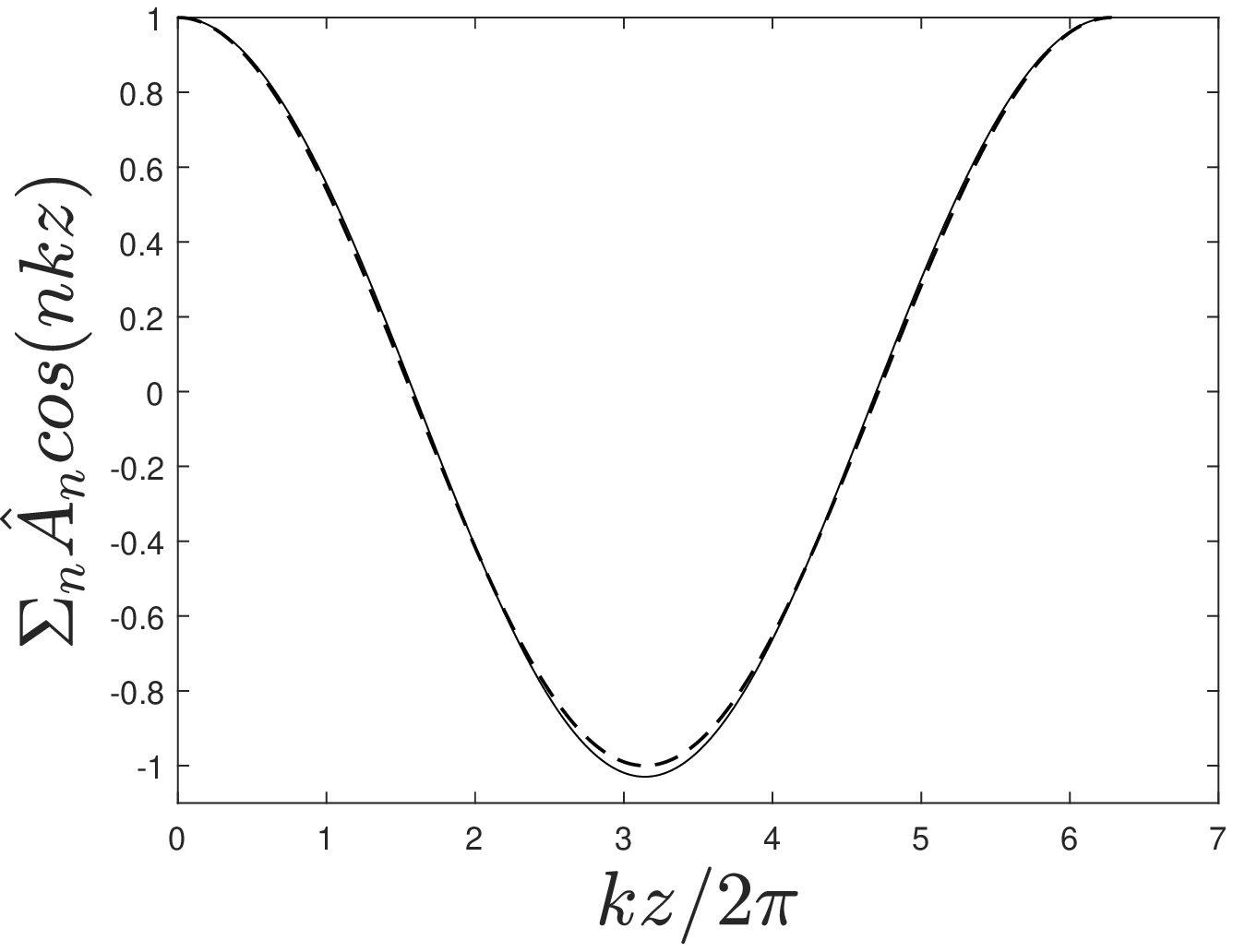}
\label{fig:str_sng_d}
}
\caption{Nonlinear chirping wave potential at $\tau = 0.0346$ for (a) the up-chirping wave with $\hat{\omega}=1.32$ and (b) the down-chirping wave with $\hat{\omega}=0.84$. The dashed curve represents the linear sinusoidal potential.}
\label{fig:str_sng}
\end{figure}

\section{Results and discussions}
\label{sec:result}

At this stage, we solve the model equations, introduced in section \ref{sec:model}, starting off the initial resonance where the system is initialised by manipulating the simulation data discussed in section \ref{sec:BOT}. As the main goal of this work, we report the results for cases which include the deepening of the wave trapping region(s) hence both particle orbit topology and particle trapping in phase-space affect the behaviour of the chirping mode simultaneously, i.e. not tractable using the flat-top model of Ref. \cite{Hezaveh2017}. This is accompanied by our observations on the behaviour of the chirping mode under the impact of multiple resonances. Accordingly, chirping waves with both downward or upward trend whose initial frequency lies in the range of magnetically passing particles are studied.

The initial plasma mode is in resonance with electrons having $1.1  \leq  \zeta_{\bf{P}} \leq 1.3$ (see  \fref{fig:eqfreq}). This corresponds to pitch-angle,
\begin{equation}
\Lambda =\frac{1}{(2\zeta_{\bf{P}}\epsilon-\epsilon+1)} ,
\label{eq:pitch}
\end{equation}
values of $ 0.65\leq \Lambda \leq 0.71$ for an inverse aspect ratio of $\epsilon=1/3$. It is noteworthy that $\Lambda=0.75$ corresponds to the trapped-passing boundary $(\zeta=1)$ in the background field. Firstly, we start the analysis by assuming that the $1^{\text{st}}$ resonance is dominant and thereby neglect higher order resonances. Subsequently, it is discussed that neglecting higher order resonances for the range of orbits under consideration is a naive assumption and there exist ranges in which a single resonance number (l) can not be regarded as the dominant one. This necessitates taking into account multiple resonances in the wave-particle interaction model. In order to demonstrate the impact of higher resonances on the nonlinear behaviour of the mode, we analyse the evolution of BGK-type chirping waves under the simultaneous influence of multiple resonances, in this case $1^{\text{st}}$ and $2^{\text{nd}}$. For each case, the validity of the adiabatic condition is analysed.

\subsection{A single resonance}
\label{sec:sngres}

We set $k_p/k_{\text{eq}}=1$ and $\zeta_{\bf{P},0}=1.176$ (see $\omega_{\text{pe}}$ on \fref{fig:eqfreq}), the self-consistent system of Eqs. \eqref{eq:sratespecial} and \eqref{eq:Fouriercoef1} is solved for both up-chirping and down-chirping modes under the impact of only the $1^{\text{st}}$ resonance and for the linear equilibrium distribution function introduced in section \ref{sec:model}. 
\begin{figure}[b]
\centering
  \includegraphics[scale=0.55]{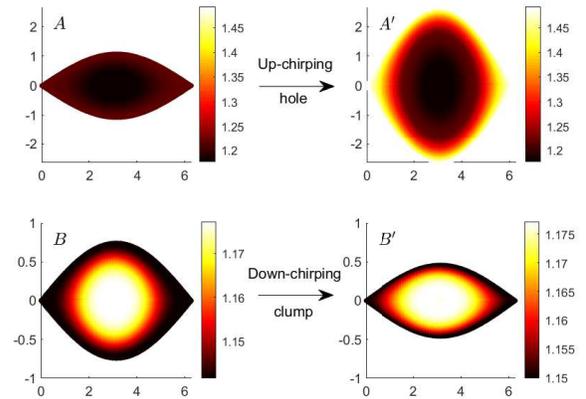}
  \caption{The phase-space islands for the single resonance case with the vertical and horizontal axis being $\tilde{J}-J_{\text{res}}$ and $\tilde{\theta}$, respectively. The color denotes the total distribution function.}
\label{fig:ps_sng}
\end{figure}
The corresponding evolution of the amplitude and the frequency is depicted in \fref{fig:evl_SR}. 
\begin{figure*}[b!]
\centering
\subfloat[]{\includegraphics[width=56mm]{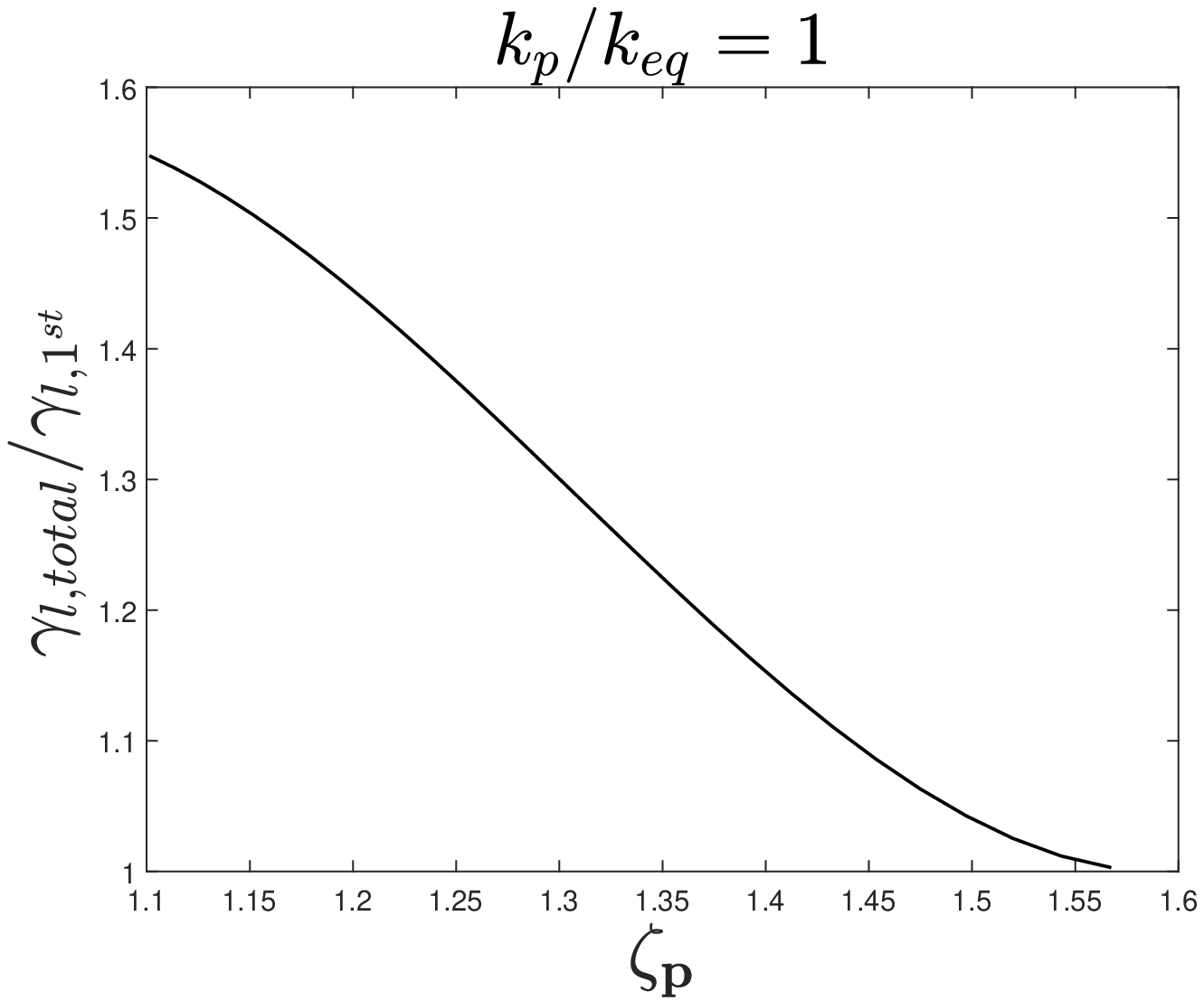}
\label{fig:k1}
}
\subfloat[]{\includegraphics[width=55mm]{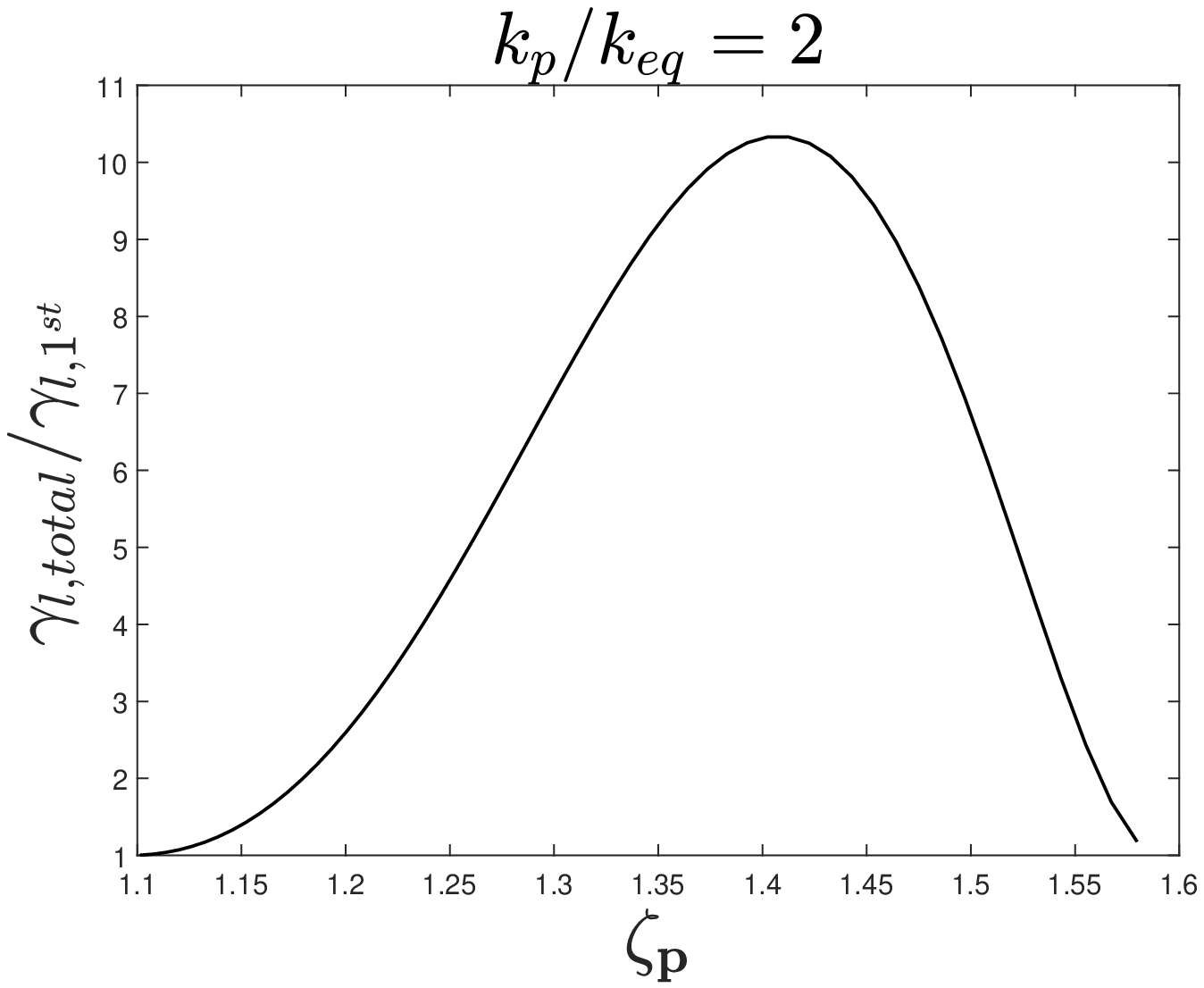}
\label{fig:k2}
}
\subfloat[]{\includegraphics[width=54mm]{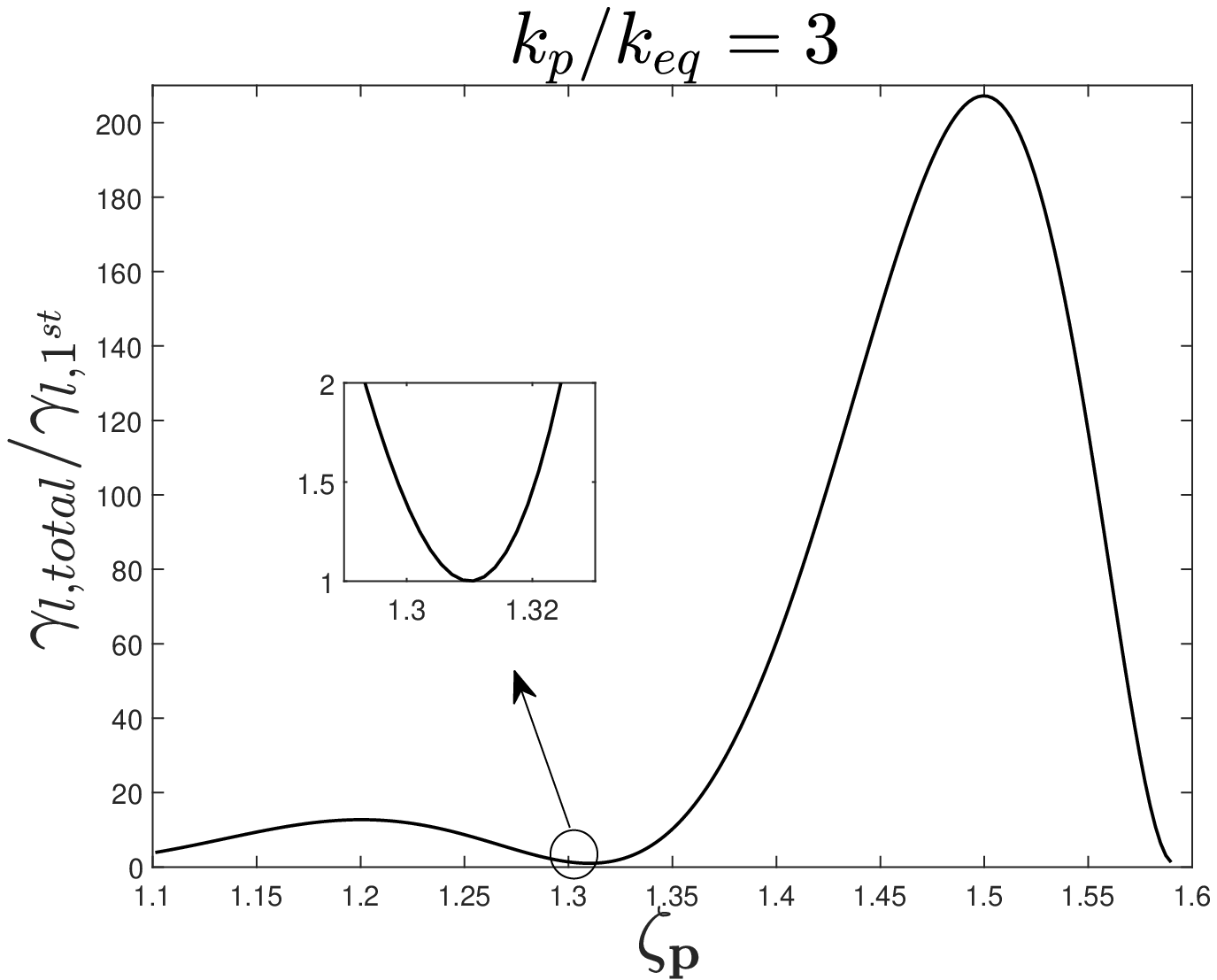}
\label{fig:k3}
}
\caption{The linear growth rate normalised to the growth rate of the $1^{\text{st}}$ resonance vs. the energy parameter as a function of $\frac{k_p}{k_{\text{eq}}}$.}
\label{fig:k}
\end{figure*}
The up-chirping mode with a growing amplitude is controlled by the dynamics of a phase-space hole $(\tilde{f}<0)$ with an expanding trapping region and therefore the effect of particle trapping in a deepening potential well is included in the behaviour of the mode. On the other hand, the downward trend is supported with a shrinking clump from which the particles are being detrapped as the mode chirps. \Fref{fig:evl_SR} shows the time-dependency of the wave parameters. The evolution of the frequency demonstrates an asymmetry in upward and downward branches. It can be observed that the upward branch is chirping faster. Both branches are initialised with the same absolute initial shift $(\abs{\delta\hat{\omega}_0}=0.055)$ in the frequency, denoted by points A and B on each curve of \fref{fig:rate_s}. The non-linear shape of the plane wave at $\tau=0.0346$ where the up-chirping and down-chirping waves experience $\approx 32\%$ and $16\%$ frequency chirping, denoted by $A^{\prime}$ and $B^{\prime}$, respectively, is shown in \fref{fig:str_sng}. This shows that the shape of the down-chirping wave is more deviated from the linear wave at this point. The phase-space density of holes and clumps is illustrated in \fref{fig:ps_sng}. Panels $A^{\prime}$ and $B^{\prime}$ illustrate the full phase-space density contours of an up-chirping hole and a down-chirping clump at $\hat{\omega}=1.32$ and $\hat{\omega}=0.84$ with  their corresponding initial separatrices shown in panels $A$ and $B$, respectively. Initial separatrices are initialised with the shape of \fref{fig:shape_hole} and $\delta \omega_0=0.055\omega_{\text{pe}}$. The particle trapping into the separatrix can be observed for the up-chirping hole, the top row, where the wave sweeps the ambient particles on its motion, as opposed to the down-chirping clump whose trapping region shrinks.  

In order to check the validity of the adiabatic condition \cite{BB1999,Eremin2002,Wang2012} given by \eqref{eq:adb_cnd}, we write it as \cite{Hezaveh2017}
\begin{equation}
\frac{\nu\gamma_{l,1^{\text{st}}}}{\omega_{\text{pe}}^2}\ll  \frac{9\pi^2\tilde{\omega}_b^2 }{16\abs{\dv{\tilde{\omega}_b}{\tau}}} ,
\label{eq:adb_cnd2}
\end{equation}
where $\tilde{\omega}_b=\frac{\omega_b}{\omega_{b,0}}$, with $\omega_{b,0}$ being the initial bounce frequency corresponding to the $1^{\text{st}}$ resonance. For a near-threshold instability with typical values of the linear growth rate to be $1$--$5$ per cent of the linear mode frequency and with $\gamma_d=0.9\gamma_l$, the LHS values lie in the range $\approx 1.8\times10^{-4}$--$0.0045$. The RHS of \eqref{eq:adb_cnd2} is shown in \fref{fig:RHS_s} as a function of time for both of the waves. This validates the adiabatic condition as the wave evolves where the RHS values increase in time.    

\subsection{Impact of higher resonances}
\label{sec:mltres}

For the range of linear plasma wave frequencies studied in Ref. \cite{Hezaveh2017}, the first resonance ($l=1$) is the dominant contributor to the interaction. In this part, we illustrate that if the $1^{\text{st}}$ resonance $(l=1)$ is formed with magnetically passing electrons having specific pitch angles, then the $2^{\text{nd}}$ resonance $(l=2)$ of the interaction can have relatively significant contribution to the wave excitation. The wave frequency studied in the previous part is an example of such cases for which  the $2^{\text{nd}}$ resonance lies in the range of magnetically trapped electrons (see \fref{fig:eqfreq}). 
\begin{figure*}[t!]
\centering
\subfloat[]{\includegraphics[width=58mm]{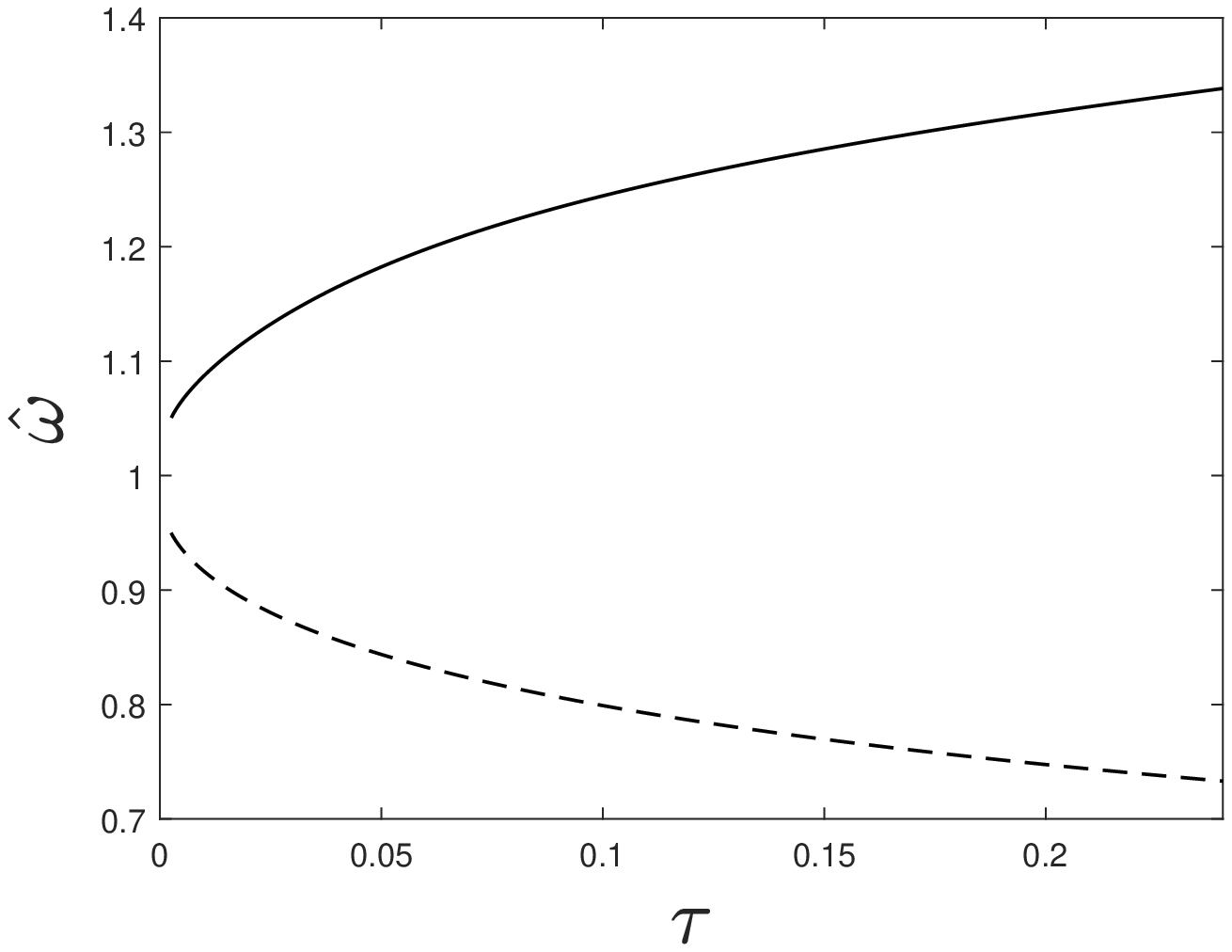}
\label{fig:}
}
\subfloat[]{\includegraphics[width=57mm]{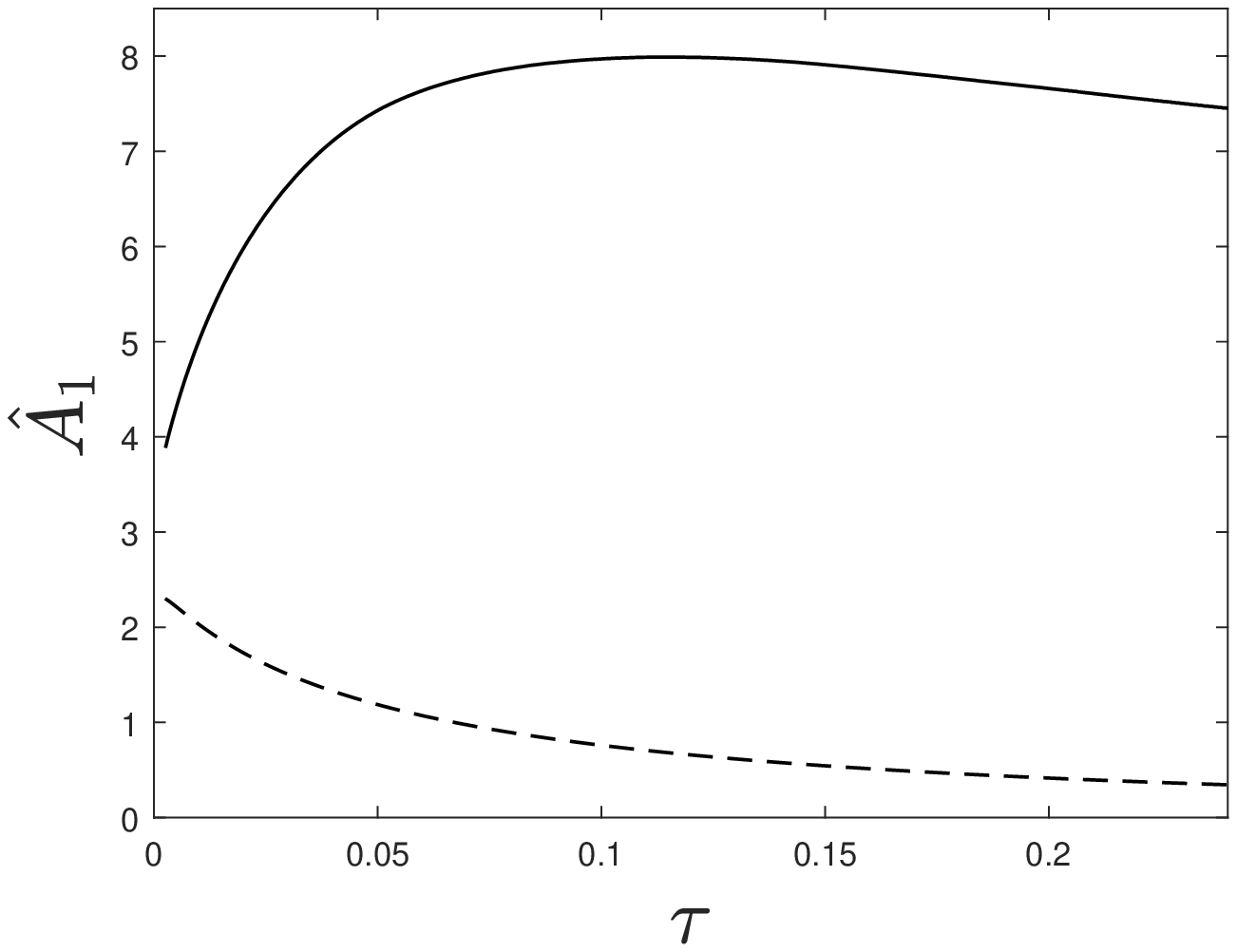}
\label{fig:A1_MR}
}
\hspace{0mm}
\subfloat[]{\includegraphics[width=56mm]{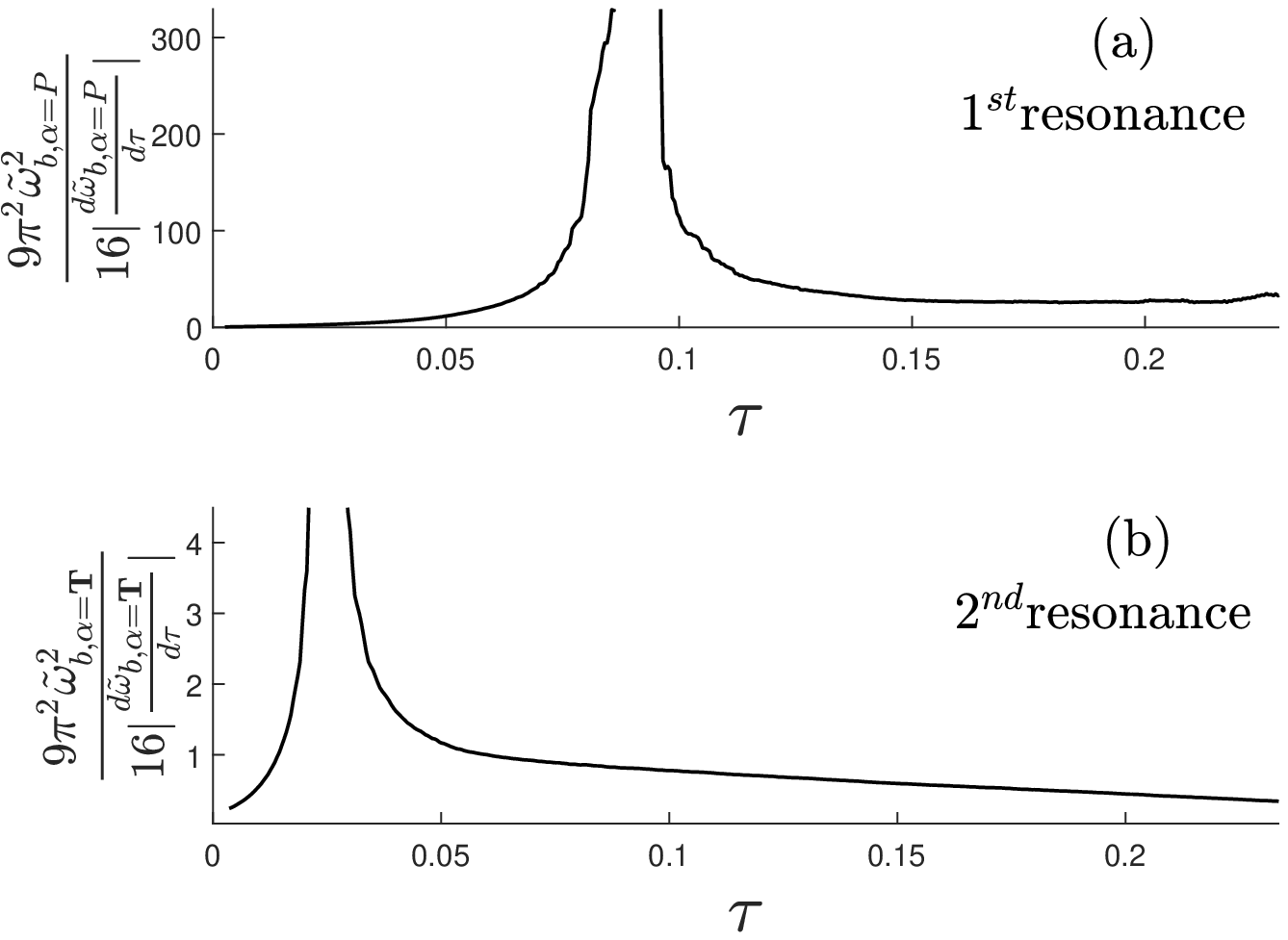}
\label{fig:RHS_MR_U}
}
\subfloat[]{\includegraphics[width=58mm]{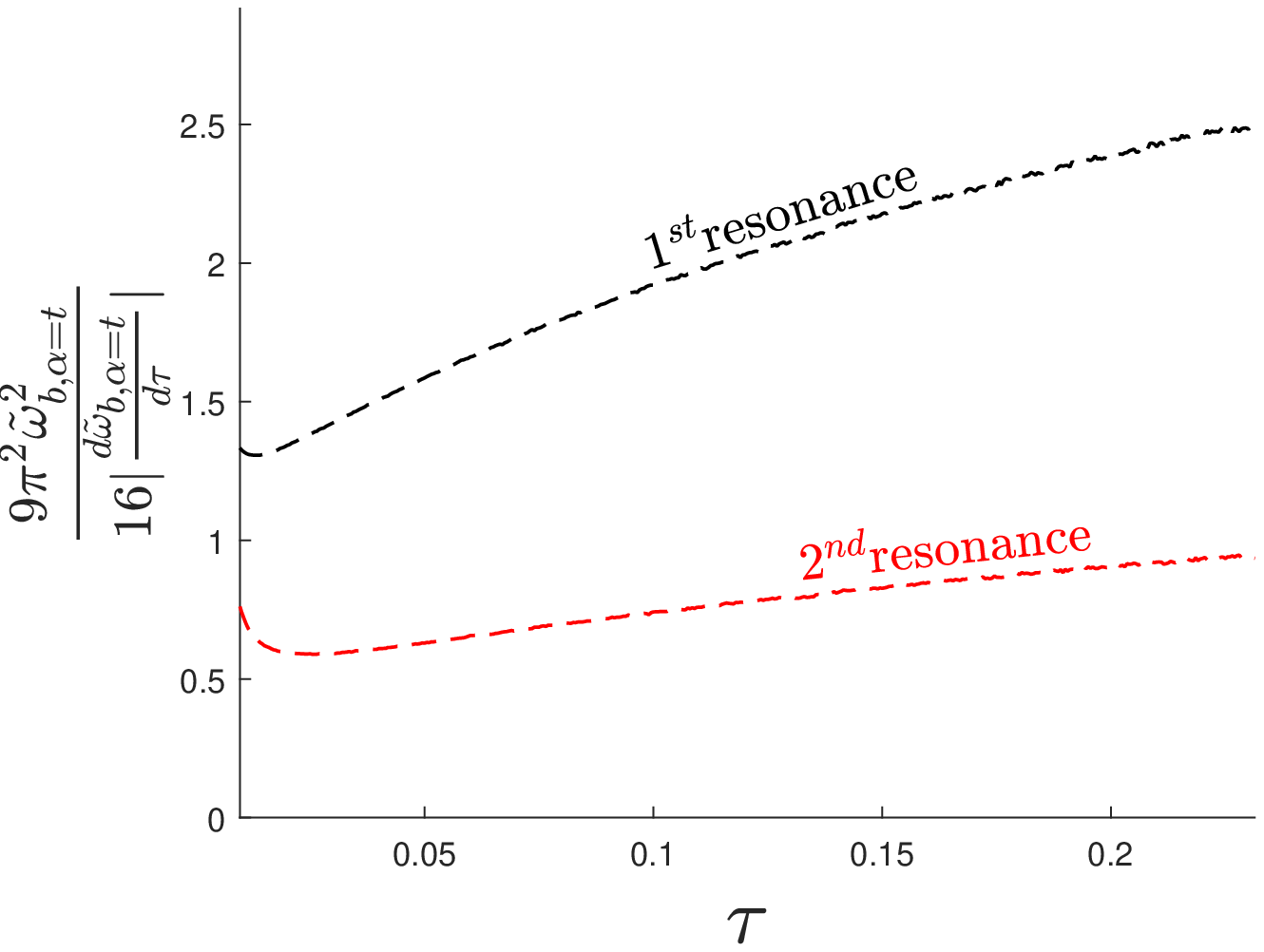}
\label{fig:RHS_MR_D}
}
\caption{Nonlinear behaviour of the chirping wave under the simultaneous effect of $1^{\text{st}}$ and $2^{\text{nd}}$ resonances. Evolution of the frequency (a), the first Fourier harmonic (b) and the RHS of the adiabatic condition at the O-point for (c) the up-chirping and (d) the down-chirping waves. In all the figures, the solid and dashed curves correspond to the up-chirping and down-chirping waves, respectively.}
\label{fig:evl_MR}
\end{figure*}
The $1^{\text{st}}$ resonance does not include any contribution from the magnetically trapped electrons since the resonance condition  with $p=1$ can not be satisfied for these electrons. However, technically speaking, the $2^{\text{nd}}$ resonance is interacting with a group of magnetically trapped electrons as well as a group of magnetically passing ones; the contribution of the latter is relatively negligible though (see section 4 and figure 8 in Ref. \cite{Hezaveh2017}). We firstly show the contribution of higher resonances by investigating their impact on the linear growth rate of the mode. For this purpose, we focus on the contribution of the $2^{\text{nd}}$ resonance and other resonances can be treated likewise. The proportion of the eigenmode wave-number ($k_{p}$) to the spatial periodicity of the equilibrium field ($k_{\text{eq}}$) is a 1D proxy for the poloidal mode numbers in realistic geometries. We denote the total growth rate, associated with the $1^{\text{st}}$ and $2^{\text{nd}}$ resonance, by $(\gamma_{l,\text{total}})$ and normalise it to the growth rate of the first resonance $(\gamma_{l,1^{\text{st}}})$. \Fref{fig:k} demonstrates $\gamma_{l,\text{total}}/\gamma_{l,1^{\text{st}}}$, where
\begin{equation}
\frac{\gamma_{l,\text{total}}}{\gamma_{l,1^{\text{st}}}} = 1 + \frac{\gamma_{l,2^{\text{nd}}}}{\gamma_{l,1^{\text{st}}}},
\label{eq:gammaltotal} 
\end{equation}
versus the energy parameter for different values of $k_{p}/k_{\text{eq}}$. It can be observed that there are regions in \fref{fig:k} where the contribution of the $2^{\text{nd}}$ resonance can be significantly higher than the $1^{\text{st}}$ one. Although these cases elaborate the significance of the higher order resonances, however, the strong dominancy of the $2^{\text{nd}}$ resonance allows neglecting the $1^{\text{st}}$ resonance and treat the interaction as having a single resonance. Accordingly, our attention is mainly focused on the more interesting regions in which one finds $1.2<\gamma_{l,\text{total}}/\gamma_{l,1^{\text{st}}}<3$, which indicates that the contribution of the $2^{\text{nd}}$ resonance is not negligible and can even be comparable to that of the $1^{\text{st}}$ resonance. Investigation of \fref{fig:k1} at $\zeta_{\bf{P},0}=1.176$ shows that the contribution of the $2^{\text{nd}}$ resonance to the interaction is more than $47\%$ of the $1^{\text{st}}$ resonance and it is not negligible. Therefore, the impact of the dynamics governed by the $2^{\text{nd}}$ resonance should be included in the analysis of the chirping waves under study. We will show the results for this choice hereafter. 

The nonlinear behaviour of the chirping waves are depicted in \fref{fig:evl_MR}. As predicted by the linear growth analysis, the inclusion of the $2^{\text{nd}}$ resonance into the interaction results in considerable change in the nonlinear behaviour of the up-chirping energetic particle driven mode. In this case, the rate of frequency chirping is smaller than the single resonance case during the evolution of the mode. The evolution of the RHS of the adiabatic condition, introduced in \eqref{eq:adb_cnd2}, is investigated at the o-point of the phase-space structures in \fref{fig:RHS_MR_U} and \fref{fig:RHS_MR_D} for the up-chirping and down-chirping waves, respectively. It is demonstrated that as the system evolves after initialisation using the BOT code data for both resonances, the value of the RHS of \eqref{eq:adb_cnd2} remains above the initial value throughout the simulation. With regards to the evolution of the first Fourier harmonic, \fref{fig:A1_MR}, the up-chirping wave initially grows faster than the single resonance case until $\tau \approx 0.009$, thereafter the rate of the amplitude change becomes smaller. Interestingly, the amplitude of the up-chirping wave saturates at $\tau \approx 0.113$ followed by a decrease. Here, we elaborate this behaviour by investigating the evolution of the phase-space structures of the up-chirping wave.

\begin{figure}[t!]
\centering
\includegraphics[scale=0.48]{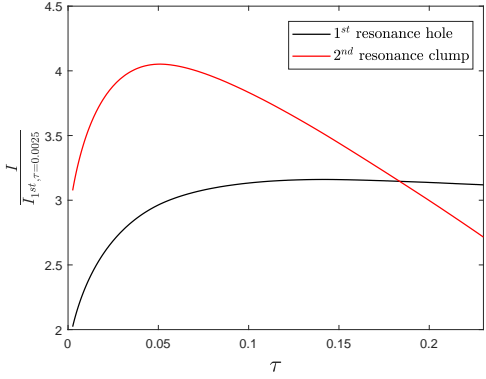}
\label{fig:ps_m_up}
\caption{Evolution of the adiabatic invariants at the separatrix for the up-chirping wave in the double-resonance case. The black and red curves correspond to the $1^{\text{st}}$ resonance hole and the $2^{\text{nd}}$ resonance clump, respectively. The y-axis values are normalised to the adiabatic invariant of the separatrix corresponding to the $1^{\text{st}}$ resonance hole at $\tau=0.0025$ denoted by $I_{1^{\text{st}},\tau=0.0025}$.}
\label{fig:adb}
\end{figure}

In the single resonance case, the phase-space structure of the up-chirping wave is a hole that constantly grows and traps ambient particles whereas for the double-resonance case, the phase-space islands supporting the up-chirping branch are a hole and a clump corresponding to the $1^{\text{st}}$ and the $2^{\text{nd}}$ resonance, respectively. The time evolution of the adiabatic invariants at the separatrix are illustrated in \fref{fig:adb} for the up-chirping wave of the double-resonance case. The values are normalised to the corresponding initial value of the $1^{\text{st}}$ resonance hole.
\begin{figure}[b!]
\centering
\includegraphics[scale=0.54]{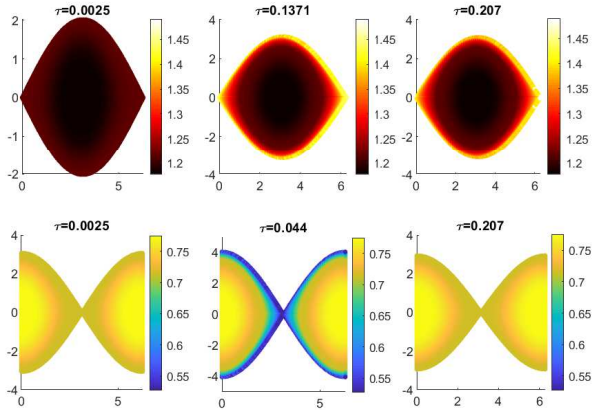}
\label{fig:ps_m_up}
\caption{Snapshots of the phase-space corresponding to the up-chirping wave with the vertical and horizontal axis being $\tilde{J}-J_{\text{res}}$ and $\tilde{\theta}$, respectively. The color denotes the total distribution function. The first and the second row correspond to the $1^{\text{st}}$ resonance hole and the $2^{\text{nd}}$ resonance clump, respectively.}
\label{fig:ps_MU}
\end{figure}
Unlike the single-resonance case, it can be observed that neither of the structures constantly grow in phase-space. The separatrix of the $2^{\text{nd}}$ resonance clump initially grows until $\tau\approx 0.051$ and then starts to shrink. This has an impact on the behaviour of the $1^{\text{st}}$ hole where it deepens until $\tau \approx0.141$. The asymptotic behaviour observed in \fref{fig:RHS_MR_U} corresponds to the times when each phase-space structure reaches the maximum expansion and does not grow further as shown in \fref{fig:adb}. At this point, the change in the bounce frequency of the trapped electrons around the O-point drops to zero (see Eq. \eqref{eq:adb_cnd2}).  

\begin{figure}[t!]
\centering
\subfloat[]{
\includegraphics[scale=0.27]{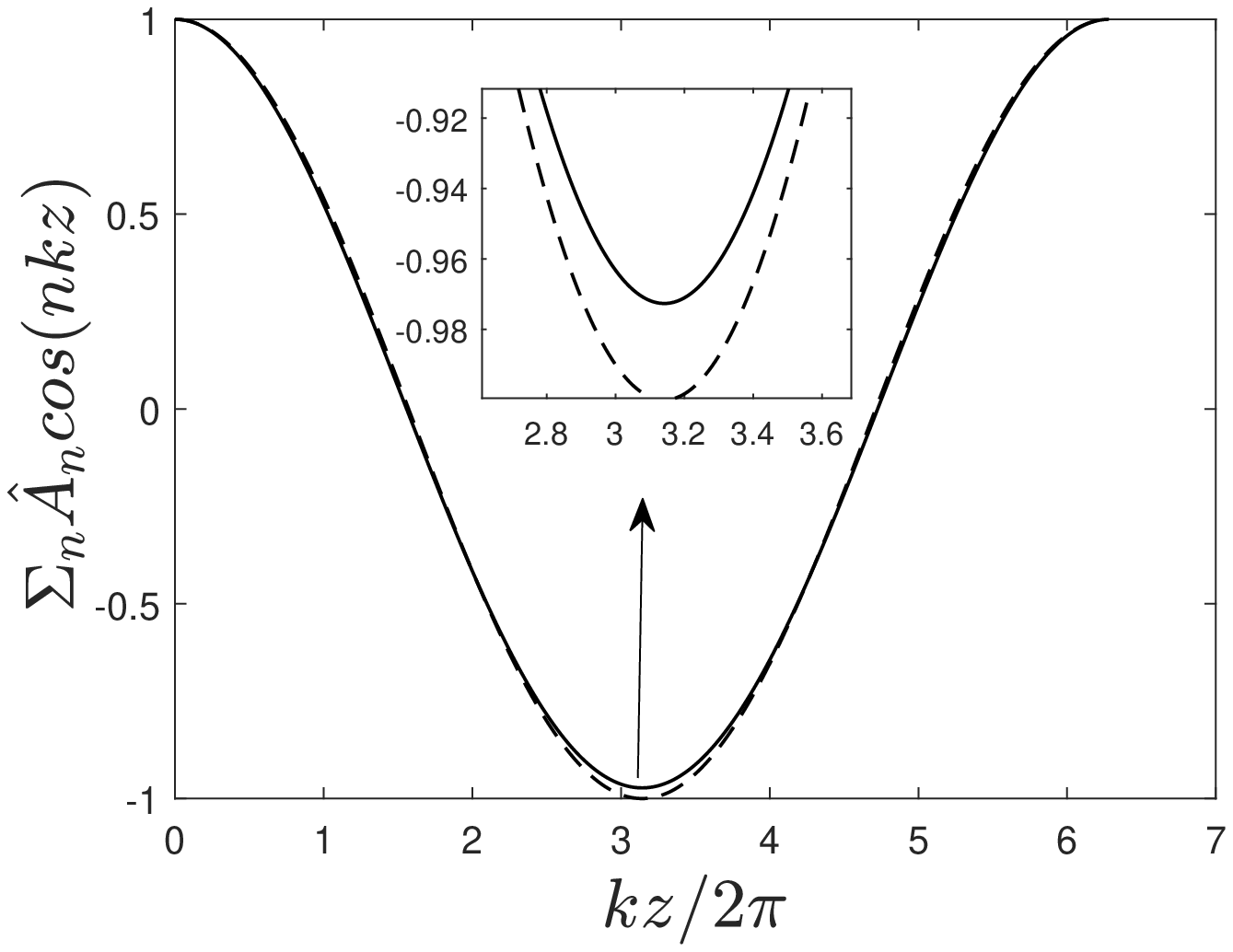}
\label{fig:str_mr_u}
}
\subfloat[]{
\includegraphics[scale=0.27]{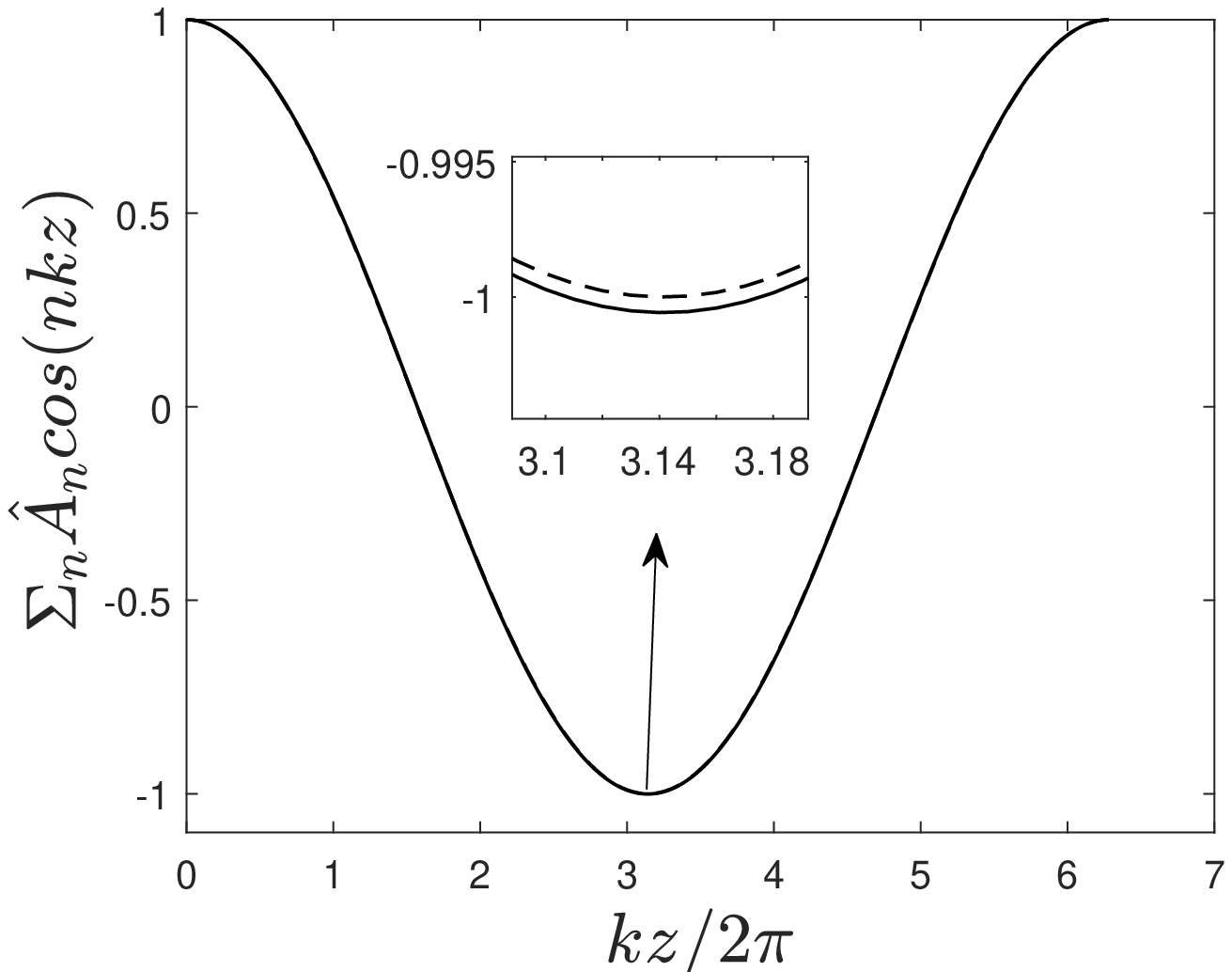}
\label{fig:str_mr_d}
}
\caption{Nonlinear potential of (a) the up-chirping wave at $\tau = 0.21$ with $\hat{\omega}=1.32$ and (b) the down-chirping wave at $\tau = 0.053$ with $\hat{\omega}=0.84$. The dashed curve represents the linear sinusoidal potential.}
\label{fig:str_M}
\end{figure}
Furthermore, snapshots of the phase-space for the up-chirping wave, illustrated in \fref{fig:ps_MU}, reveal that the particle trapping will not constantly occur as these structures evolve. As mentioned above, it can be observed that at $\tau=0.044<0.051$ the clump has expanded and trapped the ambient particles. It is worth mentioning that the $2^{\text{nd}}$ resonance clump moves towards magnetically trapped electrons having smaller values of the energy parameter $\zeta_{\bf{T}}$ and for the choice of a linear equilibrium distribution function in $\zeta$, the newly trapped electrons have smaller distribution function values. Later evolution of the clump shows a loss of the trapped particles at $\tau=0.207$. Similarly, an illustration of the phase-space hole for the  $1^{\text{st}}$ resonance at $\tau=0.1371<0.141$ shows the particle trapping inside the structure while it moves towards electrons having higher energy parameters values. At $\tau=0.207$, the separatrix of the hole has slightly shrunk as expected from \fref{fig:adb}. 

The shape of the chirping waves corresponding to the upward and downward branch is illustrated in \fref{fig:str_mr_u} and \fref{fig:str_mr_d}, respectively. The deviation of the frequency for both branches is the same as that of \fref{fig:str_sng}. Compared to the shape of the down-chirping wave, the up-chirping wave is more deviated from the linear sinusoidal wave. This is contrast with the single resonance case where the down-chirping wave experiences more change in the wave potential (see \fref{fig:evl_SR}).

\section{Summary}
\label{sec:sum}

The study of adiabatically chirping waves with deepening potentials is enabled in the trapped-passing locus model of Ref. \cite{Hezaveh2017}. This is associated with inclusion of the particle trapping effect in phase-space as the trapping region of the wave expands. This work allows the study of chirping waves with up-ward and downward frequency chirping in full range of fast particles orbit topologies which is a 1D paradigm of guiding centre motions in realistic geometries. The BOT code simulations are performed to find an appropriate shape for the phase-space structures namely holes and clumps, after their fast scale formation process. Under the adiabatic ordering, fast particles are labeled using their adiabatic invariants in a slowly evolving system. In a discretised scheme, this ordering and the Liouville theorem imply that the phase-space density remains constant in between waterbags/rings of adiabatic invariants. Hence, we resolve the perturbation of the phase-space density of fast particles using a Lagrangian mesh approach. In fact, each contour of the distribution function is considered as a waterbag. 

The evolution of the system is analysed for up-chirping and down-chirping modes in a single resonance interaction. Subsequently, we introduce regions in the fast particles orbit space where the $2^{\text{nd}}$ resonance can have remarkable contribution to the linear growth rate of the mode. This stimulates a nonlinear study of the chirping waves by including the contribution of the $2^{\text{nd}}$ resonance to the density of the fast particles. The analysis reveals that the nonlinear behaviour of the mode can be considerably altered by the $2^{\text{nd}}$ resonance. Therefore, depending on the linear frequency of the wave, it is essential to include the contribution of higher resonances when studying the evolution of chirping waves in real experiments.  

So far, in this work and the previous models on the adiabatic frequency chirping, the amplitude of the chirping wave experienced by the particles i.e. the orbit averaged mode amplitude denoted by $V$ in this work, is not a function of the phase-space action at each corresponding wave frequency. Instead, it is approximated around the centre of the separatrix by truncating the Taylor expansion of the mode amplitude around $\tilde{J}_{\text{res}}$ after the first term and is justified under the assumption of $\gamma_l \ll \omega_{\text{pe}}$, where $\gamma_l$ is the linear growth rate and $\omega_{\text{pe}}$ is the linear frequency. This implies that particle detuning from the initial linear resonance is small. However, for cases where the mode amplitude has deep gradients in the action of the fast particles equilibrium motion \cite{Revmodphys}, taking into account the higher order terms of the aforementioned Taylor expansion is a next step extension to this work which is included in our research plan.  

\section*{Acknowledgments}
This work was funded by the Australian Research Council through Grant No. DP140100790. The first author is very grateful to Dr. Robert Nyqvist for fruitful discussions that helped inspire this work.

\section*{References} 


\providecommand{\newblock}{}

\clearpage

\end{document}